\documentclass{ieeeaccess}
\usepackage{cite}
\usepackage{amsmath,amssymb,amsfonts}
\usepackage{algorithmic}
\usepackage{graphicx}
\usepackage{colortbl}
\usepackage{xcolor}
\usepackage{multirow}
\usepackage{tabularx}
\usepackage{caption}
\usepackage{textcomp}
\usepackage{wrapfig}
\usepackage{lscape}
\usepackage{longtable}
\usepackage{url}
\usepackage{subcaption}
\usepackage{ifmtarg}
\usepackage{dashrule}
\usepackage{balance}
\def\BibTeX{{\rm B\kern-.05em{\sc i\kern-.025em b}\kern-.08em
    T\kern-.1667em\lower.7ex\hbox{E}\kern-.125emX}}

\begin{document}

\title{System Identification and Controller Design for Hydraulic Actuator}
\author{\uppercase{Zainab Nisar and} 
\uppercase{Hammad Munawar}}

\address{Department of Avionics Engineering, College of Aeronautical Engineering, National University of Sciences and Technology, Islamabad, 44000, Pakistan \\ (e-mail: znisar.ms11ave@student.nust.edu.pk, h.munawar@cae.nust.edu.pk )}

\corresp{Corresponding author: Hammad Munawar (e-mail: h.munawar@cae.nust.edu.pk).}

\markboth{(IN REVIEW)}
{Nisar \MakeLowercase{\textit{et al.}}: System Identification and Controller Design for Hydraulic Actuator}

\begin{abstract}


System Identification of Hydraulic Actuators is critical for analyzing their performance and designing a suitable Control System. Hydraulic actuators are extensively used in many applications, ranging from flight simulators, robotics, orthopaedic surgery, material testing, construction and many other industrial types of machinery. In the aviation industry, hydraulic actuators are currently being used in full flight simulators used for controlling the position and orientation of the motion platform. Every actuator has its own characteristics, therefore, the choice of excitation signals for System Identification must take into account the dynamics of the actuator under consideration. This work proposes the selection of excitation signals based on bandwidth of the hydraulic actuator. Validation of the proposed selection is done by performing system identification, obtaining a mathematical model and comparing it with a nonlinear hydraulic actuator model designed in Simscape. 
After validation, a nonlinear PID control has been tuned on the identified model and tested on the nonlinear model. Extensive simulations have been run and results show accurate mathematical modelling, as well as precise control has been achieved through the proposed methodology.

\end{abstract}


\titlepgskip=-15pt

\maketitle

\section{Introduction}
\label{sec:introduction}
\PARstart Electro-hydraulic servo actuator systems (EHSAS) find significant uses in several areas due to their reliability, maintainability and high power efficiency. Consequently, they are integrated into many applications such as manipulators, vehicles, robots, and aircraft \cite{b1}. Therefore, precise control of EHSAS is an active research area and great effort has been done in the field of dynamic modelling and control of actuators to enhance performance and accuracy. 

An accurate mathematical model is required for analyzing the EHSAS dynamics and designing a control system. Mathematical model is obtained through the process of system identification which can be categorized as \textit{white box}, \textit{black box}, and \textit{grey box}. The development of white box model is entirely derived from the first principles. This kind of modelling assumes complete knowledge of the process and is determined by theoretical modelling. When enough knowledge is not available then a general model structure can be used in black box modelling. By using an estimation procedure, parameters are estimated by providing custom excitation inputs/signals to the system and measuring the output. A combination of the black and white box model is the grey box model. In addition to the knowledge from the first principles and the information contained in the measurement data, other sources of knowledge such as qualitative knowledge formulated in the rules may also be used. However, such modelling is defined by the inclusion of various kinds of easily available information \cite{b45}.
When an accurate mathematical model is available then the control scheme needs to be selected which enables precise manipulation of different performance parameters despite non-linearities, friction, noise, internal leakages, and delays. 

\begin{figure} [htb]
	\centering
	\subfloat[Flight Simulator \cite{b46}]{%
		\resizebox*{5cm}{!}{\includegraphics{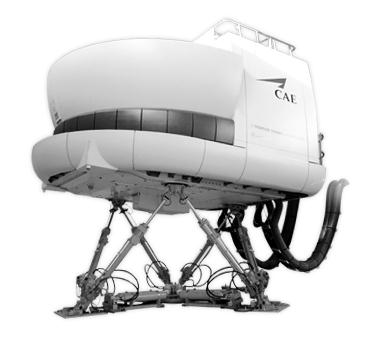}}}\hspace{5pt} 	\vspace{-8pt}
	\subfloat[Hydraulic Actuator \cite{b47}]{%
		\resizebox*{5cm}{!}{\includegraphics{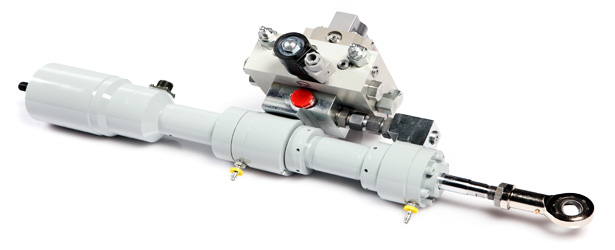}}}\hspace{5pt} 
	\caption{Example of a flight simulator with individual hydraulic actuator.} \label{sample-figure}
\end{figure}
This research aims to investigate the relation of excitation signal with the system dynamics that is further used in building the mathematical model of EHSAS with the help of system identification techniques. Although often ignored in the previous findings, this research determines the right excitation frequencies related to the system bandwidth and also defines the behavior of the system. By considering a bandwidth of the system in selection of excitation inputs, further improvements are achieved in the system identification resulting in a more accurate mathematical model. The model has been validated with the help of multiple test signals of varying amplitudes and frequencies. The proposed black-box modelling method can build and parametrize an accurate model of servo control hydraulic actuator with significant accuracy. The main contributions of this research paper are summarized below;

\begin{itemize}
	\item The Realistic Nonlinear Dynamic Model of Hydraulic Actuator of Aircraft Flight Simulator has been designed.
	\item An appropriate selection of the excitation input based on system properties has been proposed.
	\item Selection of excitation input characteristics based on system dynamics has been proposed.
	\item Experimental validation has been performed in simulation.
	
\end{itemize}

The remaining part of the paper is assembled as follows. Section II reviews existing literature in the area of system identification and control of hydraulic actuators. Section III presents the methodology portion that covers some of the basic concepts used for system identification and its different techniques and further illuminates the implementation of controller design. The paper is finally concluded in Section IV.

\section{Literature Review}
Due to their importance in industry, the control of servo-hydraulic systems has attracted a lot of attention from researchers. To obtain a mathematical model of the system, a method of identifying an actuator needs to be chosen so that an optimal accuracy of the model can be attained. This section reviews the recent literature relevant to system identification and control of hydraulic actuators. A detailed analysis of system identification covering the excitation signal, model type, model order, and validation accuracy of these techniques has been performed. The literature review is summarized in Table~\ref{Literature}.


\begin{table*} [htb]
	\scriptsize
	\centering
	\caption{Available literature on system identification and controller design techniques.}
	\vspace{1ex}
	
	\begin{tabular}{|p{0.5cm}|p{1.7cm}|p{2.0cm}|p{1.6cm}|p{1.6cm}|p{1.5cm}|p{1.7cm}|p{3.2cm}|}
	
		\hline 
	
		\textbf{ No } & \textbf{ Author } & \textbf{ Excitation Signal }  & \textbf{ Model Type }  &  \textbf{ Model Order }  &  \textbf{ Validation Accuracy}  & \textbf{ Controller Design  }  & \textbf{ Shortcomings  }    \\ \hline

		\textbf{1}    & \begin{tabular}[c]{@{}l@{}} Kalyoncu \textit{et al.}  \cite{b3} \end{tabular}   & \begin{tabular}[c]{@{}l@{}} Mathematical \\Modelling  \end{tabular} & \begin{tabular}[c]{@{}l@{}} Transfer \\Function \end{tabular} & \begin{tabular}[c]{@{}l@{}} 3rd Order \end{tabular} & \begin{tabular}[c]{@{}l@{}} ....  \end{tabular} & \begin{tabular}[c]{@{}l@{}}  Fuzzy Logic\\Controller  \end{tabular} & \begin{tabular}[c]{@{}l@{}} Model has not been acquired \\ from real system   \end{tabular} \\ \hline	
		
		\textbf{2}    & \begin{tabular}[c]{@{}l@{}} Wang \textit{et al.} \cite{ b2} \end{tabular}   & \begin{tabular}[c]{@{}l@{}} Sine Wave  \end{tabular} & \begin{tabular}[c]{@{}l@{}} State Space, \\ ARX    \end{tabular} & \begin{tabular}[c]{@{}l@{}} 3rd Order  \end{tabular} & \begin{tabular}[c]{@{}l@{}} ....  \end{tabular} & \begin{tabular}[c]{@{}l@{}} Proportion- \\Fuzzy PID \\ Hybrid  \end{tabular} & \begin{tabular}[c]{@{}l@{}} Model couldn't cover all the \\ dynamics through single \\ input frequency  \end{tabular} \\ \hline

		\textbf{3}    & \begin{tabular}[c]{@{}l@{}} Rahmat \textit{et al.} \cite{b4} \end{tabular}   & \begin{tabular}[c]{@{}l@{}} Multisine  \end{tabular} & \begin{tabular}[c]{@{}l@{}} ARX   \end{tabular} & \begin{tabular}[c]{@{}l@{}} 4th Order  \end{tabular} & \begin{tabular}[c]{@{}l@{}} 92.8\%  \end{tabular} & \begin{tabular}[c]{@{}l@{}} PID  \end{tabular} & \begin{tabular}[c]{@{}l@{}} Selection of excitation \\ signal has not been related \\ to its dynamics and \\ randomly selected its \\ frequencies  \end{tabular} \\ \hline   
		
		\textbf{4}    & \begin{tabular}[c]{@{}l@{}} Izzuddin \textit{et al.} \cite{b10} \end{tabular}   & \begin{tabular}[c]{@{}l@{}} Multisine, \\ Step  \end{tabular} & \begin{tabular}[c]{@{}l@{}} ARX  \end{tabular} & \begin{tabular}[c]{@{}l@{}} 3rd Order \end{tabular} & \begin{tabular}[c]{@{}l@{}} 95.86\% \\ 94.67\%  \end{tabular} & \begin{tabular}[c]{@{}l@{}} Predictive\\ Functional \\Control, \\ PID \end{tabular} & \begin{tabular}[c]{@{}l@{}}Selection of excitation\\ signal and their frequencies \\ have not been related \\ to its dynamics   \end{tabular} \\ \hline
		
		\textbf{5}    & \begin{tabular}[c]{@{}l@{}} Ishak \textit{et al.}  \cite{b8} \end{tabular}   & \begin{tabular}[c]{@{}l@{}} Multisine  \end{tabular} & \begin{tabular}[c]{@{}l@{}} ARX  \end{tabular} & \begin{tabular}[c]{@{}l@{}} 3rd Order  \end{tabular} & \begin{tabular}[c]{@{}l@{}} 91.16\%  \end{tabular} & \begin{tabular}[c]{@{}l@{}} Pole Placement \\ Method  \end{tabular} & \begin{tabular}[c]{@{}l@{}}  Selection of excitation\\ signal has not been related \\ to its dynamics and \\ selection on trial and \\ error base method    \end{tabular} \\ \hline
		
		\textbf{6}    & \begin{tabular}[c]{@{}l@{}} Ren G \textit{et al.} \cite{b16} \end{tabular}   & \begin{tabular}[c]{@{}l@{}} Chirp  \end{tabular} & \begin{tabular}[c]{@{}l@{}} Transfer\\ Function  \end{tabular} & \begin{tabular}[c]{@{}l@{}} 4th Order \end{tabular} & \begin{tabular}[c]{@{}l@{}} ....  \end{tabular} & \begin{tabular}[c]{@{}l@{}} Adaptive   \end{tabular} & \begin{tabular}[c]{@{}l@{}} The full capability of \\actuators has not been \\ tested   \end{tabular} \\ \hline

		\textbf{7}    & \begin{tabular}[c]{@{}l@{}} Ghazali \textit{et al.} \cite{b7}, \\ Saeed \textit{et al.} \cite{b17} \end{tabular}   & \begin{tabular}[c]{@{}l@{}} Load and Pressure \\ Values  \end{tabular} & \begin{tabular}[c]{@{}l@{}} ARX  \end{tabular} & \begin{tabular}[c]{@{}l@{}} 3rd Order \end{tabular} & \begin{tabular}[c]{@{}l@{}} ....  \end{tabular} & \begin{tabular}[c]{@{}l@{}} Hybrid Fuzzy\\ PID  \end{tabular} & \begin{tabular}[c]{@{}l@{}} Model has not been validat-\\ed on whole dynamics   \end{tabular} \\ \hline
		
		\textbf{8}    & \begin{tabular}[c]{@{}l@{}} Liang \textit{et al.} \cite{b18} \end{tabular}   & \begin{tabular}[c]{@{}l@{}} 3 Types of Signal  \end{tabular} & \begin{tabular}[c]{@{}l@{}} ARX  \end{tabular} & \begin{tabular}[c]{@{}l@{}} 3rd Order \end{tabular} & \begin{tabular}[c]{@{}l@{}} ....  \end{tabular} & \begin{tabular}[c]{@{}l@{}} Predictive   \end{tabular} & \begin{tabular}[c]{@{}l@{}} Validation has not been \\ done on their mathematical \\ model either through test \\ signals or best-fit criteria   \end{tabular} \\ \hline
		
		\rowcolor[gray]{0.75}
		\textbf{9}    & \begin{tabular}[c]{@{}l@{}}Proposed \\ Technique \end{tabular}   & \begin{tabular}[c]{@{}l@{}}Multisine, \\ Chirp \end{tabular} & \begin{tabular}[c]{@{}l@{}} Transfer \\ Function,  ARX  \end{tabular} & \begin{tabular}[c]{@{}l@{}} 3rd Order \end{tabular} & \begin{tabular}[c]{@{}l@{}} 99.17\% \\ 98.79\%  \end{tabular} & \begin{tabular}[c]{@{}l@{}} NPID   \end{tabular} & \begin{tabular}[c]{@{}l@{}} ....  \end{tabular} \\ \hline
		
	\end{tabular}
	
	\label{Literature}
\end{table*}


In 2009 Wang \textit{et al.} \cite{b2} proposed a model identification of the electro-hydraulic servo position system based on the Real-Time Workshop (RTW) hardware in the loop simulation framework, as well as the MATLAB toolbox for system identification. They obtained a 3rd order State Space and ARX model by feeding a sine wave signal as an excitation signal. The author was able to achieve very precise control through a nonlinear hybrid controller consisting of a proportional, fuzzy, and a classic PID controller. However, their system identification techniques covered a limited amount of dynamic range due to the use of single frequency.
Kalyoncu \textit{et al.} \cite{b3} suggested the Fuzzy logic controller and introduced a 3rd order transfer function. They considered the leakage flow in their mathematical modelling while most authors have ignored this phenomenon. They investigated the effects of internal leakage on the obtained mathematical model and the efficiency of the position control system even on the small spool displacement.
   
Rahmat \textit{et al.} \cite{b4} have also formulated a related work to model the electro-hydraulic actuator. The author used multisine signal and obtained the 4th order ARX model with validation of 92.8\% by using system identification techniques and further designed a PID controller for the model through simulation. The work proposed a mathematical model based on best-fit criteria, residual analysis of autocorrelation and cross-correlation. Although good results were achieved, the parameters of multisine excitation signal were selected randomly and not related to the physical characteristics of the hydraulic actuator.
Ishak \textit{et al.}  \cite{b8} designed a feedback controller with a pole placement method to achieve better system performance with more accuracy. The model was acquired by system identification based on ARX model by introducing trial and error based frequencies of multisine signal. Therefore, the actual model is close to the obtained 3rd order model with 91.16\% best-fit validation accuracy.

Izzuddin \textit{et al.}~\cite{b10} presented a system identification approach by using two types of excitation signals, the identification of the EHA system has been performed and the 3rd order ARX model was chosen. Multisine signal exhibit 95.86\% best-fit criteria whereas the step input signal showed 94.67\% fit. Other than the best-fit criteria, these models were also selected based on final prediction error (FPE) and mean square error (MSE). For position tracking, this research outlined the modelling and designing of the predictive functional control (PFC) algorithm and then compared it with the PID controller by using PSO tuning method. 
Ren G \textit{et al.}~\cite{b16} developed a controller with low bandwidth based on offline parametric linear identification technique. It was observed that the internal actuator leakage altered the model type and significantly reduced the open-loop gain that is constrained to motion. To design a controller, the mathematical model is generally obtained first by using system identification techniques. They obtained the 4th order transfer function based on the chirp input signal, but the full capability of actuator has not been tested for validation purposes.
In 2019, Liang \textit{et al.} \cite{b18} applied the technique of model predictive control (MPC) by using the optimization and constraint handling problem. This technique was validated through predictive functional control (PFC) and the results were evaluated for position control EHA system with and without disturbances in both MATLAB simulation and real-time experiments. On the other hand, the dynamic characteristics of the system gain through system identification based on 3rd order ARX model by using three different types of input signals. However, validation has not been done on their mathematical model either through test signals or best-fit criteria.  


Li \textit{et al.} \cite{b5} introduced a methodology by an integrating black and grey box model identification for deriving mathematical model of an electro-hydraulic servo system. A white box model of the entire system was build and the unknown parameters were calculated approximately as per prior experience.  In this research, to enhance the model of a white box, the identification of the black box model was applied to change certain parameters of the model. Finally, the identification of the grey box model has been done using the improved model as a new initial model. Therefore, simulation results and their comparison with measurement data showed an accurate model of the system.
Ghazali \textit{et al.} \cite{b7} dealt with system identification using recursive or offline techniques and proposed a 3rd order ARX model by varying load and pressure values. As the significant effect in model parameters have been improved, despite that, the model has not been validated on whole dynamics.
In 2019, Liyang \textit{et al.} \cite{ b15} presented a double layered network scheme identification, a combination of black and grey box method. However, a precise model is obtained by adding the two layers separately. Based on the validation tests, the obtained model is highly compatible with the actual model. 

  
Moreover, disturbance, stability, and speed are some dynamics of the system, that were being used to improve the system by adding controllers in the existing system.  In this regard, Zhong proposed an algorithm \cite{b9} that is the combination of fuzzy logic and neural network techniques, as a means to suppress the non-linearities and disturbances for such systems.
Wonohadidjojo \textit{et al.} proposed a Fuzzy logic controller to overcome the non-linearities and Particle Swarm Optimization (PSO) method used in his research to obtain the best value for tuning its parameters \cite{b6}.
Saeed \textit{et al.} \cite{b17} introduced a hybrid fuzzy PID tracking methodology for the electro-hydraulic servo system from the perspective of heavy manufacturing processes. The aim was to build a nonlinear hybrid controller comprising of a classical PID controller, fuzzy logic controller, and a fuzzy-PID controller based on the self-adjusting modifying factor that significantly improves the robustness of the system and its dynamic and static properties. Researchers have introduced a variety of other controllers as well to overcome the non-linearities of the system, such as linear, nonlinear controller, and artificial intelligence approaches such as PID \cite{b11}, model predictive control (MPC) \cite{b12}, sliding mode control (SMC) \cite{b13}, and adaptive control \cite{b14}. 

After reviewing the state-of-the-art relevant literature, it has been determined that a methodology for relating excitation signals with dynamics of the system to be identified, has not been presented. Therefore, in this research, the primary focus is to present a framework for the selection of correct parameters of excitation signals. Our work improves upon the existing research in this field by characterizing the excitation signal in relation to the dynamics of the system to be identified. As a first step, a realistic nonlinear dynamic model of EHSAS has been developed in Simulink. The effects of compressibility, friction, internal servo valve leakage, actuator leakage, and inertia have been included to make the model more accurate. Once the mathematical model is obtained through system identification, a nonlinear PID controller has been used to implement the precise position control of an electro-hydraulic servo system.

\section{Methodology and Experimental Validation}
Firstly, the following section explains the dynamics of the system. Then based on dynamics, the nonlinear hydraulic actuator system has been modelled by using Simscape physical system toolbox where all components of the hydraulic actuator have been provided, which are particularly applied to control the displacement of the hydraulic actuator. Furthermore, system identification techniques have been done to acquire an accurate model and for that, a suitable selection of excitation signal and model have been identified. Consequently, after the EHSAS model as shown in Fig. \ref{basicflow} is acquired, both the simulation and the position control of EHSAS experiments have been performed based on the obtained model. Moreover, a nonlinear PID controller has been implemented on the nonlinear model and the identified model. 

\subsection{Dynamics of the Plant}
The EHSAS to be studied here in this paper is a \textit{Moog MCR-M-1002} hydraulic actuator that is commonly used in the motion platform of flight simulators. The plant is also composed of a servo amplifier, servo valve by \textit{Moog 725-106}, and the load. The whole system is operated by a hydraulic pump equipped with safety valves. To make the system more efficient, the system also incorporates the following sensors: 1 linear variable differential transducer (LVDT) that governs the position of each cylinder, two pressure transducers; one for piston side and second for rod side chamber pressure, and two limit switches to interpret the feedback of full extension and retraction of the cylinder. 

\begin{figure}[t]
	\centering
	\resizebox{8.7cm}{!}{\includegraphics{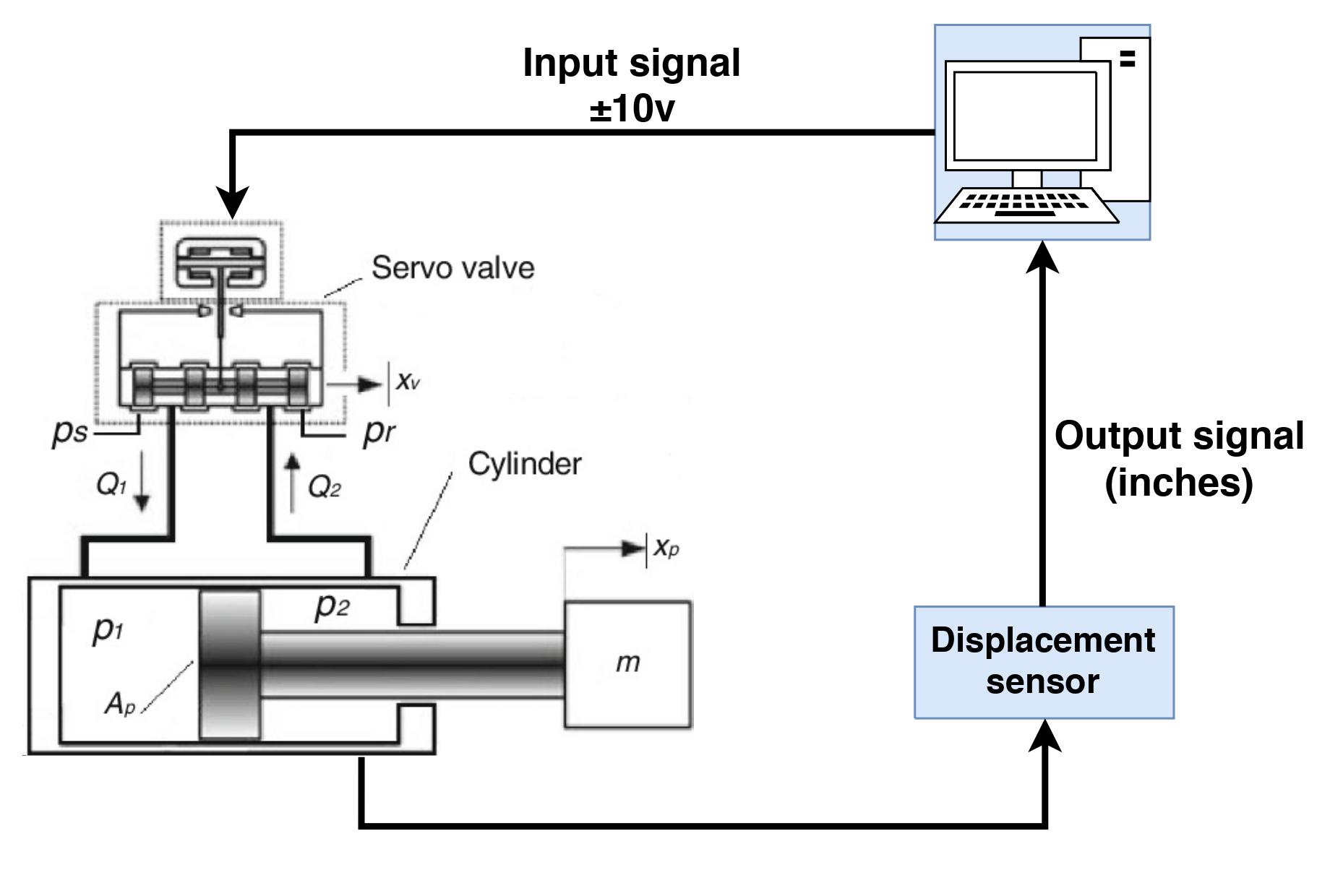}}	
	\caption{Basic flow of EHSAS.}
	\label{basicflow}
\end{figure}

In this study, the EHSAS under consideration consists of two major components: the valve and the cylinder. The cylinder has been modelled along with a double actuator with load mounted at the end of the rod. The servo valve and actuator are depicted in the Fig. \ref{basicflow}. The control signal is represented by $u$, the displacement of the cylinder is represented by $X_p$, $A_p$ is the area of hydraulic cylinder, the fluid flow to and from the cylinder is $Q_1$ and $Q_2$, respectively. The fluid pressure within chamber 1 of the cylinder is $P_1$ and the chamber 2 of the cylinder is $P_2$. On both sides, the pressurized areas are $A_1$ and $A_2$, respectively. When there is a pressure difference between $P_1$ and $P_2$ the cylinder will extend or retract on its position. The servo valve controls the fluid flow $Q$ in each chamber. The basic fluid flow is expressed by Equation~\ref{equ1 flow}.

\begin{equation}
Q = K_v X_v \sqrt{\Delta P_v}
\label{equ1 flow}
\end{equation}
where $K_v$ is servo valve gain, $X_v$ denotes spool valve displacement that is controlled by input signal $u$ and $P_v$ is the pressure difference. The fluid flow towards the cylinder and back to the cylinder is represented by Equation~\ref{equ2 fluid flow} and ~\ref{equ3 fluid flow}~\cite{b40}. 

\begin{equation}
Q_1 =
\begin{cases}

K_1 X_v \sqrt{P_s - P_1}\ ; \: \: \: \: \: \: \: \: \: \: X_v \ge 0 \\
K_1 X_v \sqrt{P_1 - P_r}\ ; \: \: \: \: \: \: \: \: \: \: X_v < 0  \\
\end{cases}
\label{equ2 fluid flow}
\end{equation}

\begin{equation}
Q_2 =
\begin{cases}
-K_2 X_v \sqrt{P_2 - P_r}\ ; \: \: \: \: \: \: \:  X_v \ge 0  \\
-K_2 X_v \sqrt{P_s - P_2}\ ; \: \: \: \: \: \: \:  X_v < 0  \\
\end{cases}
\label{equ3 fluid flow}
\end{equation}
\\
Therefore, pressure can be obtained for each chamber by specifying the relation between the bulk modulus, volume, and the flow rate. The fluid pressure $P_1$ and $P_2$ can be written as:

\begin{equation}
P_1 = \frac{\beta}{V_{1}} \int(Q_1 - \frac{dv_1}{dt})dt  \\     
\end{equation}

\begin{equation}
P_2 = \frac{\beta}{V_{2}} \int(\frac{dv_2}{dt} -Q_2 )dt \\
\end{equation}
\\
Now, the Equation~\ref{valve eq} presents a relation between the spool valve displacement $X_v$, and the input voltage signal $u$, where the servo valve gain $K_v$ acts as the constant of proportionality.

\begin{equation}     
X_v = K_v u  \\
\label{valve eq}
\end{equation}    
Hence, the hydraulic system dynamics for fluid flow $Q_{L}$ has been obtained from a Tailor Series Linearization.

\begin{equation}
Q_{L} = K_{q} X_v - K_{c} P_L 
\label{load flow eq}
\end{equation}
Because of the fluid flow $Q_{L}$, the load pressure $P_L$ can be defined as the pressure across the servo actuator and the first derivative equation of the load pressure is defined by the division of the total flow through the fluid capacitance and hydraulic actuator. Here, $K_{q}$ represents flow gain coefficient and $K_{c}$ shows the flow pressure coefficient. So, the first derivative of the load pressure in the form of mathematical equation is given by~\cite{b50}.

\begin{equation}
\dot{P_{L}} = \dfrac{4\beta}{V_{t}} (Q_L - C_{tp} P_L - A_p \dot{X_p}) 
\label{load pressure eq}
\end{equation}

Here, $\beta$ defines the bulk modulus, $C_{tp}$ represents the total leakage coefficient, and $V_t$ is total volume of the fluid. Then, after adding and substituting the above equations, the displacement in terms of derivative becomes:

\begin{equation}
\dddot{X_p} = A^{2}_{p}\dfrac{4\beta}{V_{t} M} \dfrac{K_q K_v}{A_p} u - \dfrac{4\beta}{V_{t}} (K_c + C_{tp}) \ddot{X_p} - A^{2}_{p}\dfrac{4\beta}{V_{t} M} \dot{X_p} 
\label{diff eq}
\end{equation}

By rearranging the above equations, we get:

\begin{equation}
\dddot{X_p} + \dfrac{4\beta}{V_{t}} (K_c + C_{tp}) \ddot{X_p} + A^{2}_{p}\dfrac{4\beta}{V_{t} M} \dot{X_p}  = A^{2}_{p}\dfrac{4\beta}{V_{t} M} \dfrac{K_q K_v}{A_p} u 
\label{diff eq2}
\end{equation}

From Equation~\ref{diff eq2}, the continuous time transfer function is given as:

\begin{equation}
\dfrac{X_p(s)}{U(s)} = \dfrac{A^{2}_{p}\dfrac{4\beta}{V_{t} M} \dfrac{K_q K_v}{A_p} } {s^{3} +s^{2} \dfrac{4\beta}{V_{t}} (K_c + C_{tp}) + s A^{2}_{p}\dfrac{4\beta}{V_{t} M}  }
\label{final_eq}
\end{equation}

Therefore, transfer function equation~\ref{final_eq} can be written in terms of constant values that are:

\begin{equation}
\dfrac{X_p(s)}{U(s)} = \dfrac{\omega_{a} K_{a} }{ s^{3} + 2 \zeta_a \omega_{a} s^{2} + \omega_{a} s }
\label{final_eq2}
\end{equation}

In equation~\ref{final_eq2}, actuator gain $K_a$, natural frequency $\omega_a$, and servo valve damping coefficient $\zeta_a$ can be defined as:

\begin{align*}
K_a &=\dfrac{K_q K_v}{A_p} \\ \\
\omega_a &= A_p \sqrt{\dfrac{4\beta}{V_{t} M}} \\ \\
\zeta_a &= \dfrac{\sqrt{\dfrac{4\beta}{V_{t}} (K_c + C_{tp})}}{2A_p}
\end{align*}
 
Therefore, after reviewing the generic mathematical model in Equation~\ref{final_eq2}, it has been proposed that the transfer function of EHSAS can be suitably presented as a third order transfer function. The basic structure of the proposed continuous time transfer function is as shown in Equation~\ref{basic eq}.

\begin{equation}
G(s)=\dfrac{X_p(s)}{U(s)} = \dfrac{b_1 s^{2} + b_2 s + b_3}{ s^{3} + a_1 s^{2} + a_2 s + a_3}
\label{basic eq}
\end{equation}

\begin{figure}[t]
	\centering
	\includegraphics[height=11cm,width=0.5\textwidth]{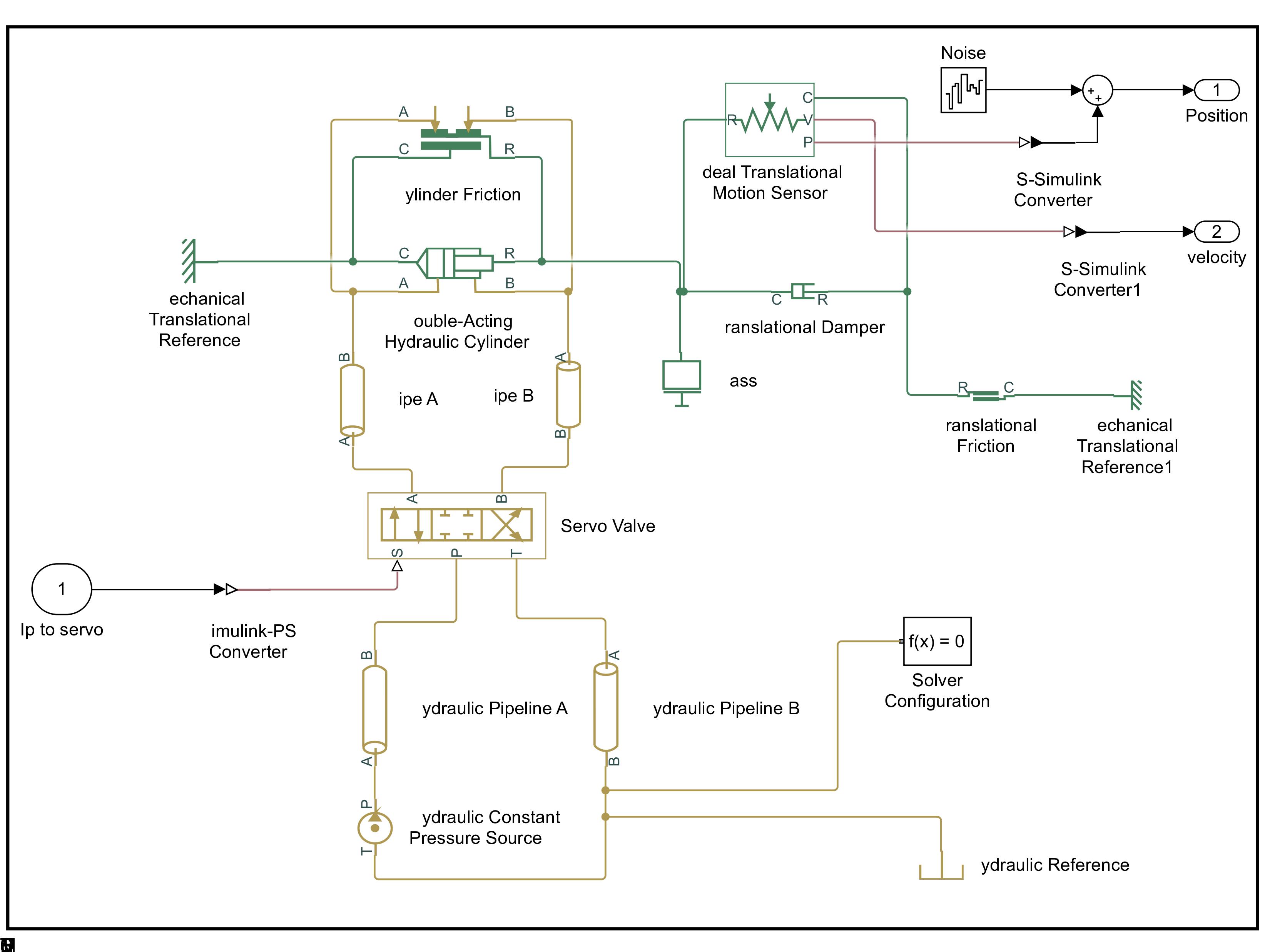}
	\caption{Nonlinear model of EHSAS.}
	\label{simscape model}
\end{figure} 

\subsection{Nonlinear modelling based on actual parameters}
This section discusses the dynamical modelling of the hydraulic servo system. The system has been modelled as a nonlinear system in Simscape. After such parameter identification, the nonlinear model has been modified into a mathematical form in which the physical structure is represented in a nonlinear generalized model. To develop a mathematical model of the system, different dynamic parameters have been used. Each of the Simscape model components have been configured with parameters of the real hydraulic actuator taken from its technical manuals and also shown in Table \ref{parameters}. This nonlinear model as shown in Fig. \ref{simscape model} has been generated as a replacement of the real actuator, so that system identification and experimentation techniques would be applied without any damage to the hydraulic actuator.

\begin{center}
	\begin{table} [t]
		\centering
		\caption{Parameters of EHSAS \cite{b48}.}
		\vspace{1ex}
		\label{parameters}
		
		\begin{tabular}{|p{1.3cm}|p{3.2cm}|p{0.8cm}|p{1.5cm}|}
			
			\hline 
			\textbf{Components}   & \textbf{Parameters} & \textbf{Symbols}  & \textbf{Values} \\
			\hline
			 & \begin{tabular}[c]{@{}l@{}} Input Signal\end{tabular}  & \begin{tabular}[c]{@{}l@{}} $u$ \end{tabular} & \begin{tabular}[c]{@{}l@{}} $\pm $ 10 $V$ \end{tabular}   \\
							
			\begin{tabular}[c]{@{}l@{}} Servo Valve  \end{tabular} & \begin{tabular}[c]{@{}l@{}}Flow Discharging Coefficient \end{tabular} & \begin{tabular}[c]{@{}l@{}} $C_f$ \end{tabular} & \begin{tabular}[c]{@{}l@{}}0.6 \end{tabular} \\
						
			\begin{tabular}[c]{@{}l@{}} (\textit{Moog} \end{tabular} & \begin{tabular}[c]{@{}l@{}} Leakage Area \end{tabular} & \begin{tabular}[c]{@{}l@{}} $k_a$ \end{tabular} & \begin{tabular}[c]{@{}l@{}} 1e-12 $m^{2}$ \end{tabular} \\
						
			\begin{tabular}[c]{@{}l@{}} \textit{725-106}) \end{tabular} & \begin{tabular}[c]{@{}l@{}} Maximum Opening \end{tabular} & \begin{tabular}[c]{@{}l@{}} $O_m$ \end{tabular} & \begin{tabular}[c]{@{}l@{}} 0.0178 $m$  \end{tabular} \\
						
			& \begin{tabular}[c]{@{}l@{}} Servo Valve Area \end{tabular} & \begin{tabular}[c]{@{}l@{}} $A_s$ \end{tabular} & \begin{tabular}[c]{@{}l@{}} 0.0002318 $m^{2}$  \end{tabular}\\
						
			& \begin{tabular}[c]{@{}l@{}} Servo Valve Gain \end{tabular} & \begin{tabular}[c]{@{}l@{}} $K_v$ \end{tabular} & \begin{tabular}[c]{@{}l@{}} 2.2e-6 $m/V$ \end{tabular} \\ \hline
		
			 &  \begin{tabular}[c]{@{}l@{}} Piston Stroke  \end{tabular} & \begin{tabular}[c]{@{}l@{}}  $X_s$ \end{tabular} & \begin{tabular}[c]{@{}l@{}} 60 $in$ \end{tabular} \\
			
		 	& \begin{tabular}[c]{@{}l@{}} Contact Stiffness \end{tabular} & \begin{tabular}[c]{@{}l@{}} $C_s$ \end{tabular} & \begin{tabular}[c]{@{}l@{}} 6.14e8 $N/m$ \end{tabular} \\
						
		\begin{tabular}[c]{@{}l@{}} Hydraulic \end{tabular}	& \begin{tabular}[c]{@{}l@{}} Contact Damping \end{tabular} & \begin{tabular}[c]{@{}l@{}} $C_d$ \end{tabular} & \begin{tabular}[c]{@{}l@{}} 200 $N/(m/s)$ \end{tabular}\\
						
		\begin{tabular}[c]{@{}l@{}} Actuator  \end{tabular}	& \begin{tabular}[c]{@{}l@{}} Piston Area \end{tabular} & \begin{tabular}[c]{@{}l@{}} $A_p$ \end{tabular} & \begin{tabular}[c]{@{}l@{}} 12.5 $in^{2}$ \end{tabular} \\
						
		\begin{tabular}[c]{@{}l@{}} (\textit{Moog MCR-}  \end{tabular}	& \begin{tabular}[c]{@{}l@{}} Flow Gain Coefficient \end{tabular} & \begin{tabular}[c]{@{}l@{}} $K_q$ \end{tabular} & \begin{tabular}[c]{@{}l@{}} 1.8e-6 $m/V$ \end{tabular}  \\
						
		\begin{tabular}[c]{@{}l@{}}  \textit{M-1002}) \end{tabular}	& \begin{tabular}[c]{@{}l@{}} Actuator Gain \end{tabular} & \begin{tabular}[c]{@{}l@{}} $K_a$\end{tabular}  & \begin{tabular}[c]{@{}l@{}} 491.04e-12 \end{tabular} \\
						
			& \begin{tabular}[c]{@{}l@{}} Dead Volume  \end{tabular} & \begin{tabular}[c]{@{}l@{}} $V_d$  \end{tabular} & \begin{tabular}[c]{@{}l@{}} 0.0003048 $m^{3}$ \end{tabular} \\
						
			& \begin{tabular}[c]{@{}l@{}} Specific Heat Ratio \end{tabular} & \begin{tabular}[c]{@{}l@{}} $h$  \end{tabular} & \begin{tabular}[c]{@{}l@{}} 1.4 \end{tabular} \\
						
			& \begin{tabular}[c]{@{}l@{}} Bulk Modulus  \end{tabular} & \begin{tabular}[c]{@{}l@{}} $\beta$  \end{tabular} & \begin{tabular}[c]{@{}l@{}} 22e4 $psi$ \end{tabular} \\
						
			&  \begin{tabular}[c]{@{}l@{}} Load \end{tabular} & \begin{tabular}[c]{@{}l@{}} $M$ \end{tabular} & \begin{tabular}[c]{@{}l@{}} 500 $kg$ \end{tabular} \\
						
			
			& \begin{tabular}[c]{@{}l@{}} Damping Coefficient \end{tabular} & \begin{tabular}[c]{@{}l@{}} $B_s$ \end{tabular} &  \begin{tabular}[c]{@{}l@{}} 100 $N/(m/s)$ \end{tabular} \\
			
			& \begin{tabular}[c]{@{}l@{}} Spring Stiffness \end{tabular} & \begin{tabular}[c]{@{}l@{}} $K_s$ \end{tabular} &  \begin{tabular}[c]{@{}l@{}} 20 $Nm$ \end{tabular} \\
			
			& \begin{tabular}[c]{@{}l@{}} Coulomb Friction Force \end{tabular} & \begin{tabular}[c]{@{}l@{}} $\alpha_1$ \end{tabular} & \begin{tabular}[c]{@{}l@{}} 450 $N$  \end{tabular} \\
						
			& \begin{tabular}[c]{@{}l@{}} Viscous Friction Coefficient \end{tabular} & \begin{tabular}[c]{@{}l@{}} $\alpha_2$ \end{tabular} &  \begin{tabular}[c]{@{}l@{}} 64 $N/(m/s)$ \end{tabular} \\

			\hline
		\end{tabular}
	\end{table}
\end{center}

\subsection{System Identification}
Although hydraulic actuator can be fully defined by its physical laws as shown in Subsection $A$. However, with repeated usage, the system develops leaks or the performance of its components deteriorates and many uncertainties occur in the system. System identification techniques use measured experimental data to obtain frequency responses with the intent that, system uncertainties are observed and the models obtained are more precise.  

\subsubsection{Selection of Input Excitation Signal}
Based on the dynamics of the system and the requirements of modelling, researchers have presented various excitation signals. The most commonly used excitation signal is Pseudo-Random Binary Sequences (PRBS) which is periodic and deterministic signal with the replacement of noise. An integer number of periods in PRBS should be used to enjoy its good properties that limit the choice of the length of experiment \cite{b19}. For the identification of linear systems, PRBS signal is widely used. Therefore, it cannot be used for nonlinear systems as they also require the judicious choice of excitation amplitude to cover the entire operating range of the system to identify.

Another commonly used excitation signal is a step signal, that can give the different response parameters which are very helpful in system identification. However, the step signal does not cover the full dynamics of the system as it contain low information content in data \cite{b49}. Therefore, it is not considered for this research. Sinusoidal is another useful signal, which covers the given frequency with different possible amplitudes. However, to cover the full dynamic range of the system, the sinusoidal signal is given as a chirp or multisine signal so that more frequencies can be captured. After extensive literature review of system identification of similar hydraulic actuators, it has been determined that the most suitable excitation signals for black-box identification of EHSAS are multisine and chirp signal as presented in Table~\ref{Literature}. 

Chirp excitation signal is a frequency sweep signal, with increasing frequencies exponentially over a certain period. Chirp signal has the same crest factor as sinusoid and the excited frequency band is well controlled. The chirp excitation signal is designed according to the \cite{b44}, as shown in Equation~\ref{equ_chirp}.

\begin{equation}
X_{chirp}(t) =Acos(\phi(t)) 
\label{equ_chirp}
\end{equation}
In equation \ref{equ_chirp}, $A$ represents the amplitude of the signal and $\phi(t)$ shows the phase of signal.

\begin{align*}
\phi(t) &= \phi_0(t) +2\pi( \dfrac{k}{2} t^{2} + f_{0}(t) ) \\
k &= \dfrac{f_1 - f_0}{T}
\end{align*}

$f_{0}$ is the starting frequency at $t=0$, $f_{1}$ is final frequency and $T$ is the time that is taken to sweep from starting to final frequency.

\subsubsection{Selection of  Parameters of Excitation Signal}
Formation of the selected excitation signals requires the determination of amplitude and frequency. However, the reviewed literature has used hit and trial to select the excitation signal parameters and then perform experimental validation to check the required accuracy of mathematical model has been achieved. Therefore, in this research the parameters of excitation signals have been proposed according to the actuator dynamics. A reference has been obtained from system identification of complex nonlinear systems such as aircraft and rotor-craft \cite{b20}.
There is no need for higher frequency inputs. In order to obtain a good identification model for flight mechanics and control applications, the maximum chirp signal frequency for aircraft and rotor-craft is limited to almost 2 Hz. For forming a chirp signal, the minimum $\omega_{min}$ and maximum $\omega_{max}$ frequencies are required to ensure that the excitation signal covers the full range of dynamics of the system. However the excitation signal must also be related to dynamics of the system. Therefore, a range of frequency has been proposed with relation to the bandwidth of the system. The range of frequency for chirp signal is
$0.1\omega_{Bw} \leq \omega \leq 2\omega_{Bw}$. 	
Finally, the maximum amplitude has been selected which avoids the actuator saturation to excite the system and to improve signal-to-noise ratio (SNR) \cite{b20}. 

Similarly, multisine excitation signal has also been used for capturing the whole dynamics of the system. The combination of three different frequencies with the relation to its system behavior has been proposed instead of distribution over a range of frequencies. Equation~\ref{equ_multi} shows the proposed multisine equation with selected parameters.

\begin{multline}
X_{multisine}(t) = A[sin[0.1 \omega_{Bw}(t)] + sin[0.5 \omega_{Bw}(t)] + \\
sin[2 \omega_{Bw}(t)]]
\label{equ_multi}
\end{multline}

\begin{figure}[t]
	\centering
	\begin{subfigure} [b]{0.5\textwidth}
		\includegraphics[width=1\textwidth]{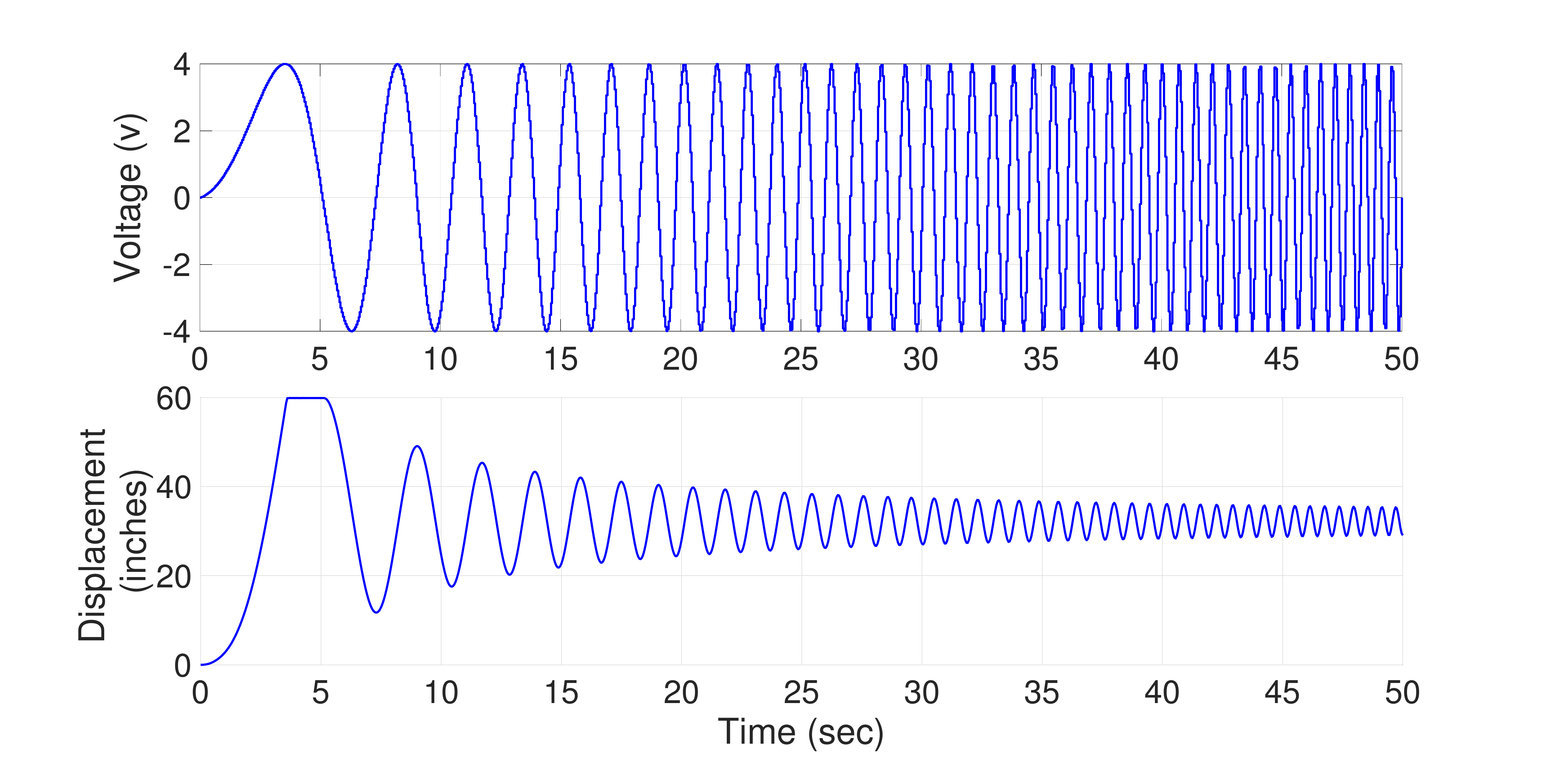}  
		\caption{Chirp Excitation Signal}
		\label{chirp_data}
	\end{subfigure}
	\begin{subfigure} [b]{0.5\textwidth}
		\includegraphics[width=1\textwidth]{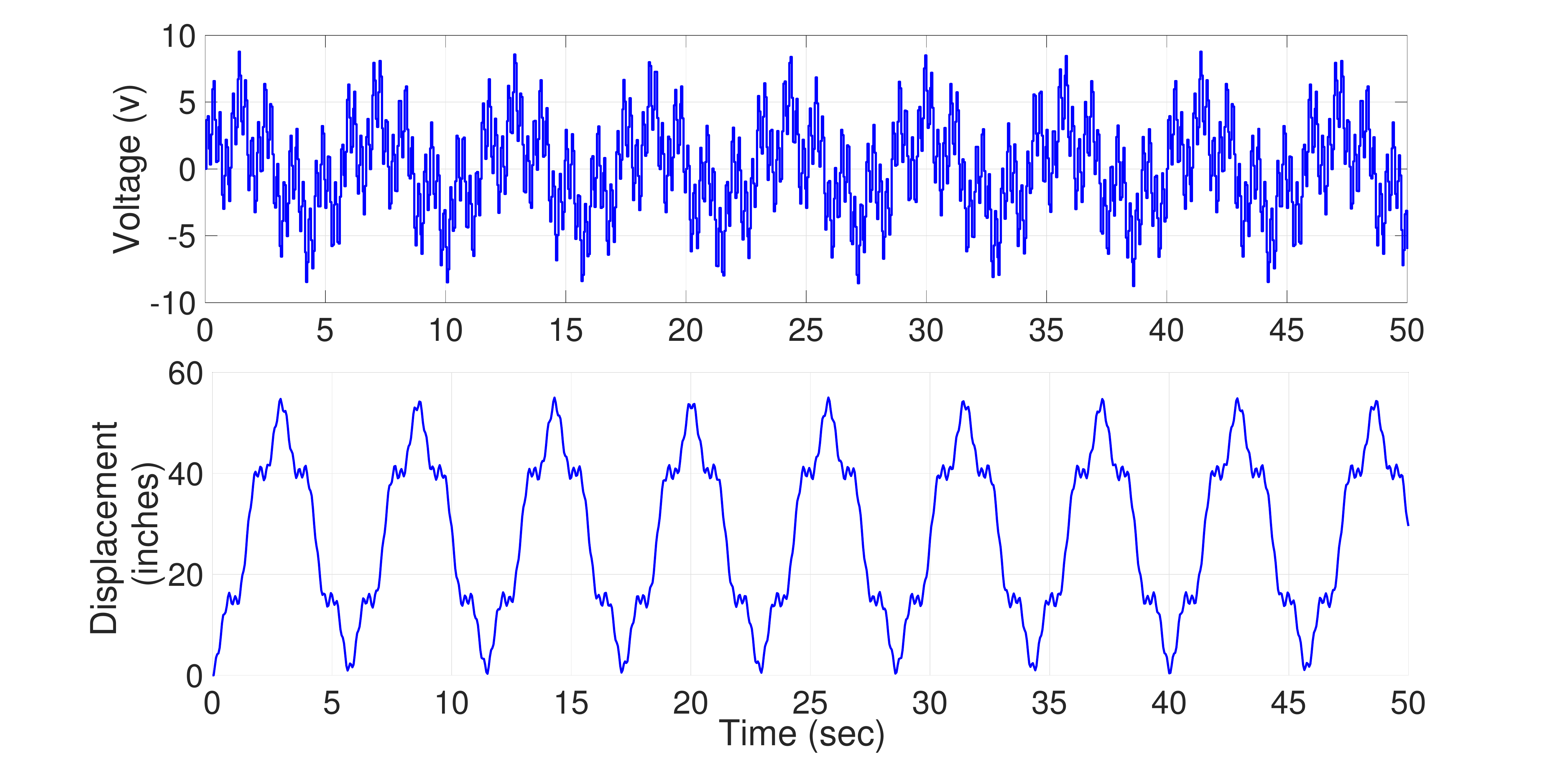}  
		\caption{Multisine Excitation Signal}
		\label{mutisine_data}
	\end{subfigure}
	
	\caption{Data collection for model identification.}
\end{figure}

\subsubsection{Model Identification}
Linear model is used as the identification of EHSAS to approximate a nonlinear model. Since this is the discrete time model that can describe the relationship between $u(t)$ and $X_p(t)$. The linear model is preferred over nonlinear model as it is more popular model among different estimation methods to identify the EHSAS, while at the same time, it has the ability to represent the real system with high accuracy~\cite{b23}. Different researchers have successfully modelled the hydraulic actuator as a linear system. Therefore, a 3rd order transfer function model and ARX model have been preferred that will result in a 3rd order transfer function, as discussed in Section~\ref{Literature}. To estimate the unknown parameters of the system based on the ARX model, the least square method is used. Furthermore, it also has been experimentally validated.

The first step in the identification of model is to get system input and output signals. The chirp excitation signal and the corresponding output as shown in Fig. \ref{chirp_data}, that is divided into two sets within 50s; $80\%$ set is used for estimation and the remaining $20\%$ is used for validation. 

Model ARX-331 obtained from one of the sets of data as a result of highest best-fit value with 50ms sampling time, that is known as $G_{chirp}(s)$.

Multisine signal is the second type of excitation signal used to get the EHSAS model. Fig. \ref{mutisine_data} displays the estimation and validation obtained data. The first half of the obtained data used for estimation and second for validation.
The 3rd order transfer function, $G_{multisine}(s)$ has been chosen to represent the multisine excitation model since it has the highest best-fit value. 
\begin{figure}[t]
	\centering
	\begin{subfigure} [b]{0.5\textwidth}
		\includegraphics[width=1\textwidth]{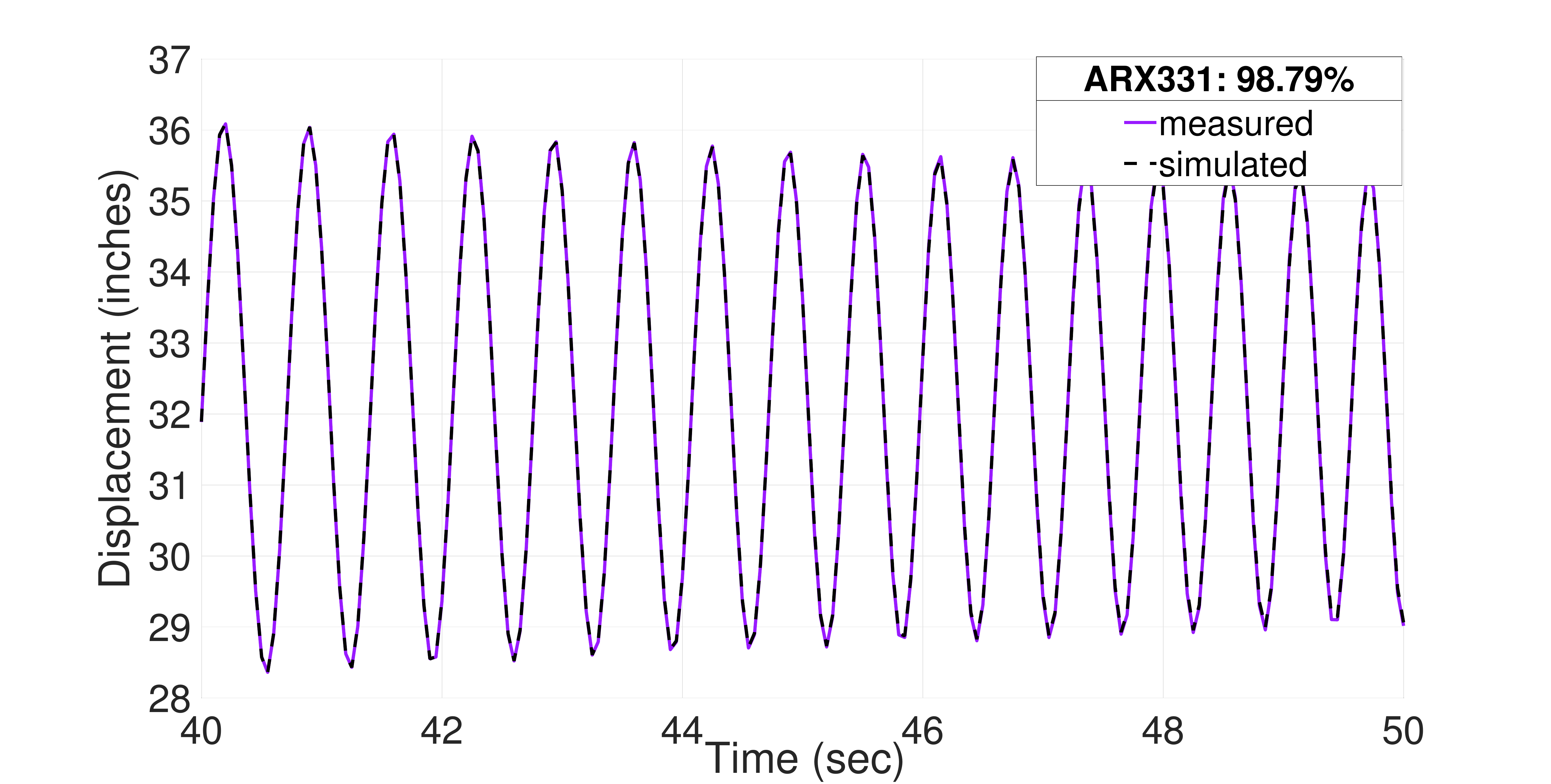}  
		\caption{Chirp Excitation Signal}
		\label{validation_chirp}
	\end{subfigure}
	\begin{subfigure} [b]{0.5\textwidth}
		\includegraphics[width=1\textwidth]{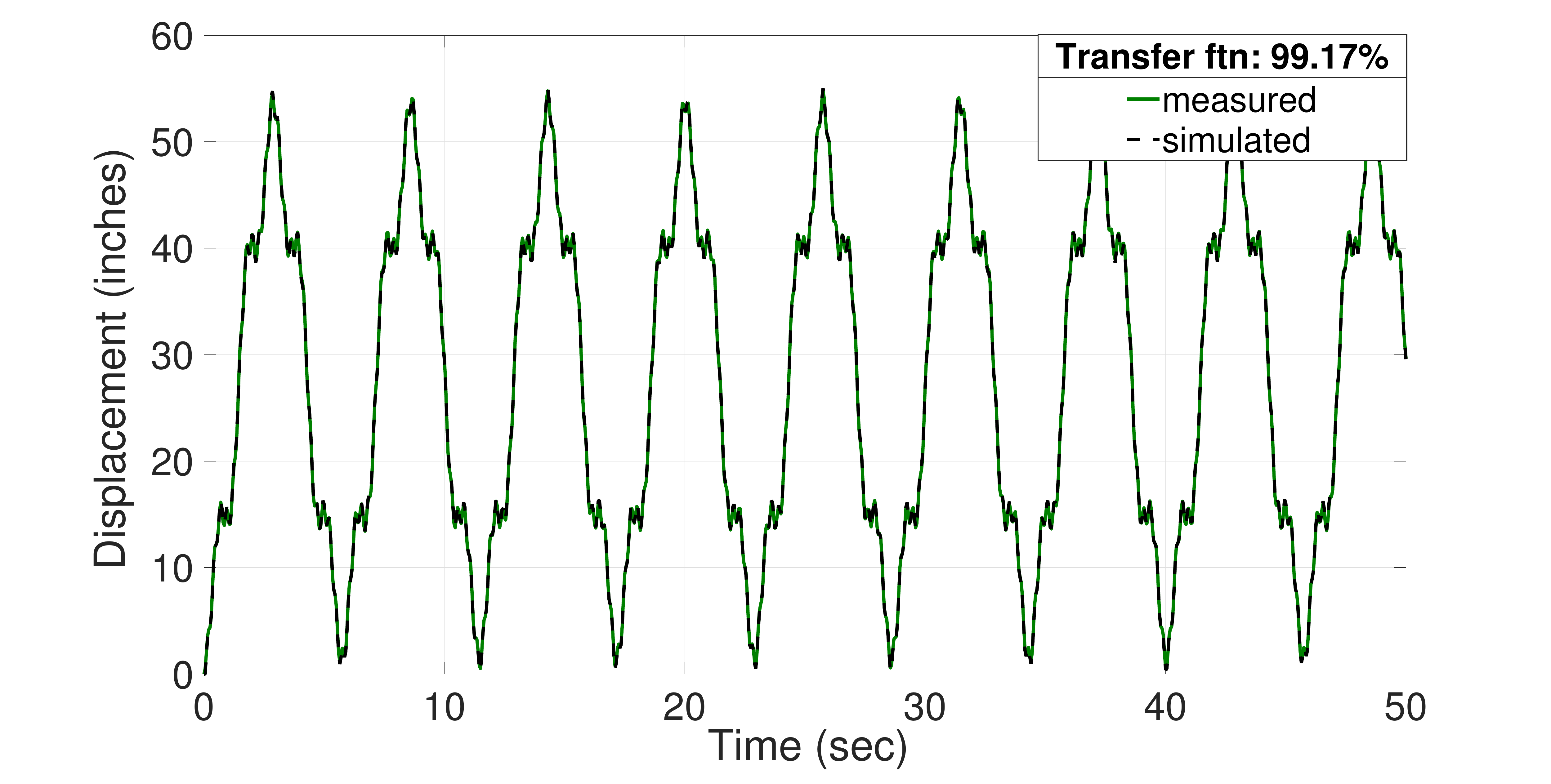}  
		\caption{Multisine Excitation Signal}
		\label{validation_multisine}
	\end{subfigure}
	\caption{Measured and simulated output data.}
	\label{validation_data}
\end{figure}

\subsubsection{Model Validation}
The model validation has been done by evaluating the best-fit between the measured and simulated data. The model with a higher percentage of best-fit indicates that the model can represent a model close to the nonlinear model. In addition to the best-fit, mean square error (MSE) and final prediction error (FPE) has been observed for model identification as tabulated in Table \ref{error_selection}. In Fig. \ref{validation_data}, it is shown that the model obtained through multisine and chirp excitation signals provides an accurate measured and simulated output.


\begin{center}
	\begin{table}[htb]
		\centering
		\centering\caption{Model selection criteria.}
		\vspace{1ex}
		\label{error_selection}
		\begin{tabular}{|l|l|l|}
			
			\hline
			\textbf{Model Parmeters} \: \: \: \: \: \:  & \textbf{Multisine} \: \: \: \: &\textbf{Chirp} \: \: \: \: \\                                  \hline
			\textbf{Best-Fit Percentage} & 99.17\%     & 98.79\%  \\ \hline
			\textbf{MSE}  & 0.002918  & 0.00301 \\                \hline
			\textbf{FPE}  & 0.00304  & 0.00311 \\                        \hline
			
		\end{tabular}
	\end{table}
\end{center}

\begin{figure} [t]
	\centering
	\subfloat[Chirp Model]{%
		\resizebox*{8cm}{!}{\includegraphics{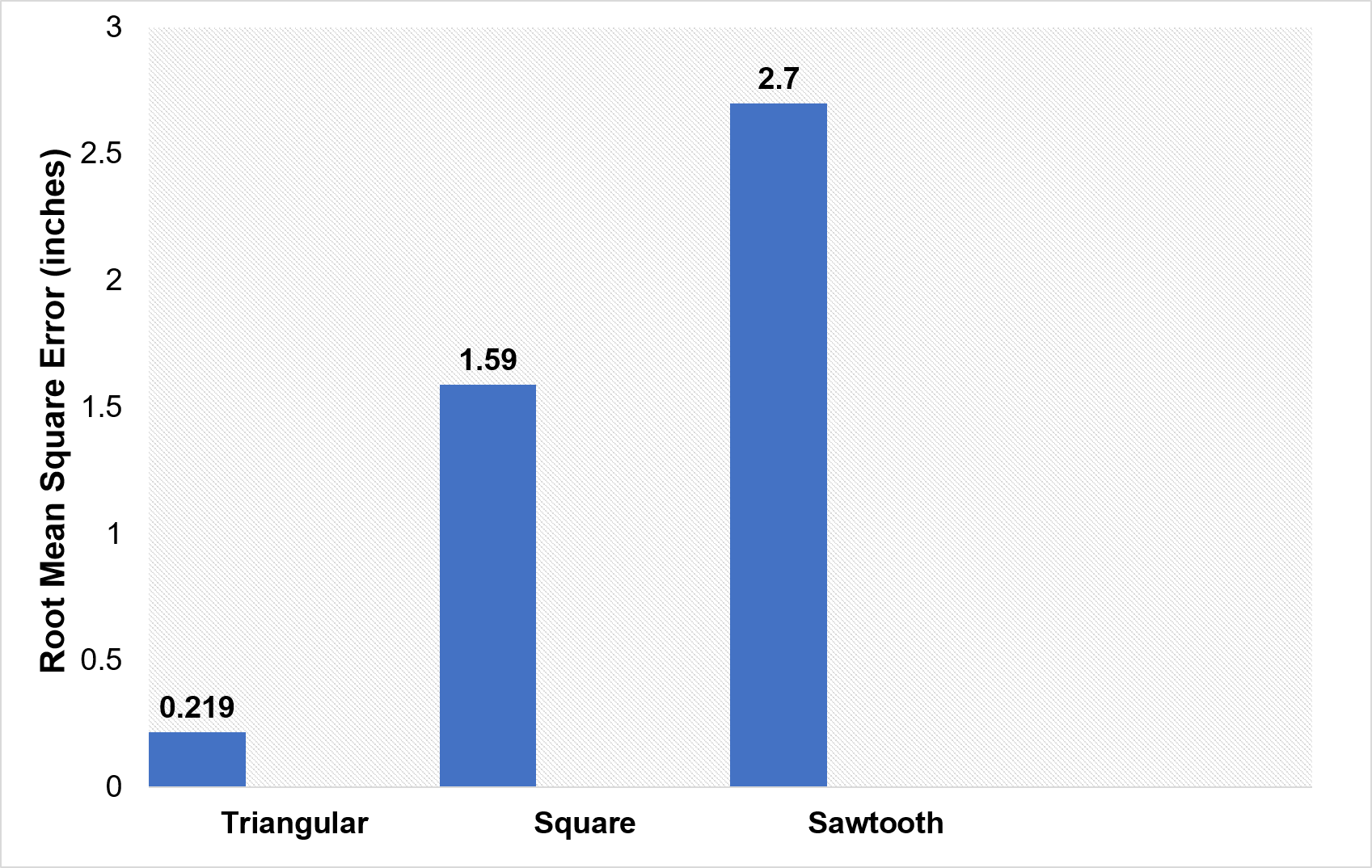}}}\hspace{5pt} 	\vspace{-8pt}
	\subfloat[Multisine Model]{%
		\resizebox*{8cm}{!}{\includegraphics{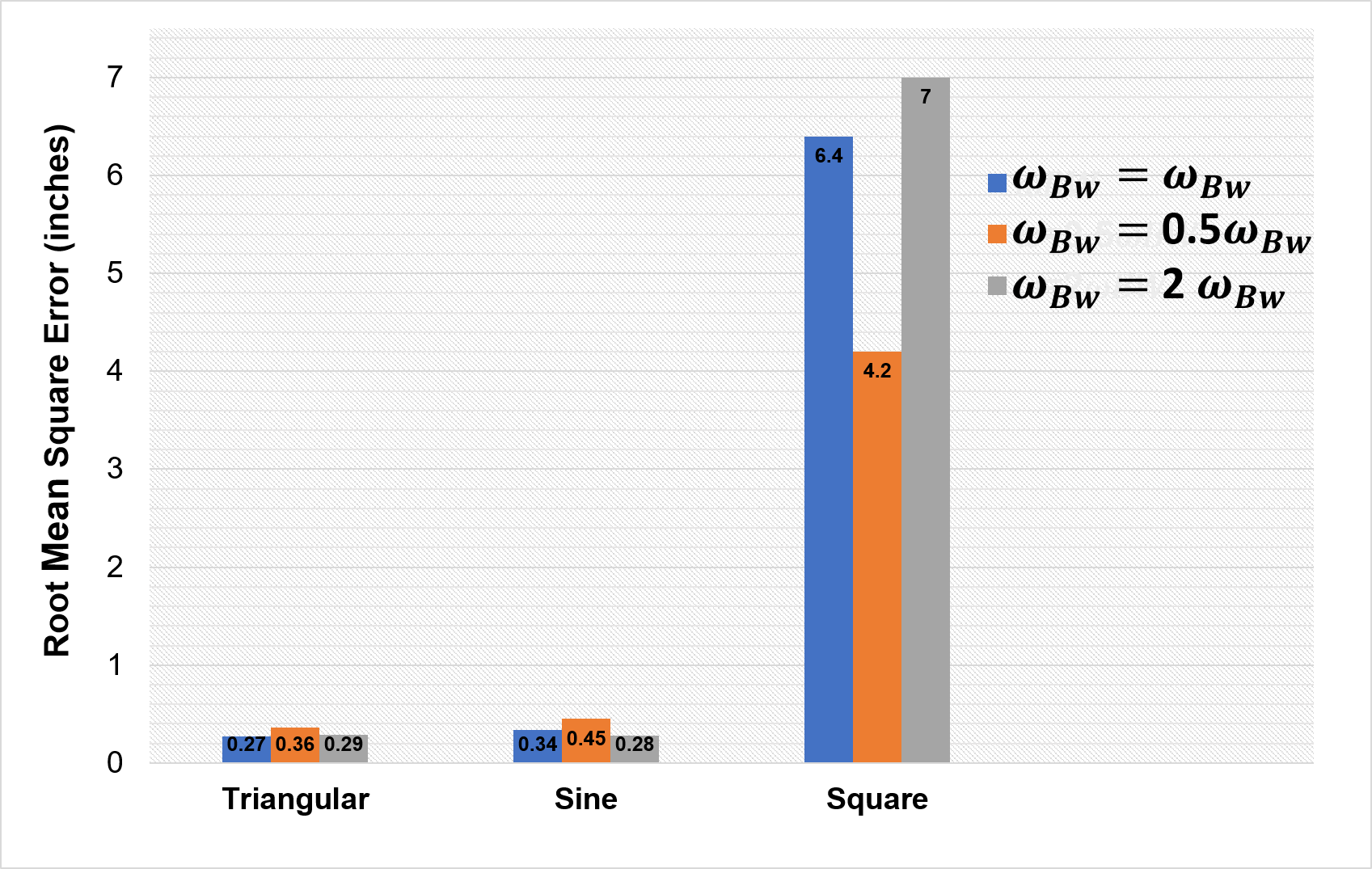}}}\hspace{5pt} 
	\caption{Experimental analysis through RMSE.} 
	\label{validation_bw}
\end{figure}

Besides the best-fit percentage, FPE and MSE of multisine and chirp excitation signal, the validation has been done. For the evaluation of identified model, different test signals including triangular, square, sine, and sawtooth have been fed to the model of chirp and multisine. Therefore, in order to draw a comparison between nonlinear model and identified model, the root mean square error (RMSE) has been calculated as shown in Fig.~\ref{validation_bw}, respectively. It is clearly depicted that the lowest error calculated by chirp excitation signal responded to different test signals are triangular, followed by square signal and sawtooth signal. Similarly, in multisine excitation signal by varying amplitude and frequency of test signals, it has been noted that there is the minimum error obtained from triangular test signal then sine signal followed by square signal. Thus, if the comparison has been drawn between chirp and multisine excitation signals then it has been claimed that chirp excitation signal shows better performance with minimum RMSE.



\subsection{Controller Design}
After the system identification has been completed with a reasonable accuracy, the nonlinear PID controller has been selected for control. The most widely used control scheme in various applications is PID. The PID controller has been used very effectively for better position tracking performance, and fast response time~\cite{b30}. Since, it has basic structure having different types tunning methods available that are more efficient as well. By tuning the PID gain $K_p$, $K_i$ and $K_d$ values, system performance such as rise time, overshoot, settling time, and steady state error can be significantly improved. The conventional type of controller is often difficult to acquire optimum performance because of the presence of non-linearities in the system. As the dynamic model of the hydraulic actuator system incorporates major non-linearities. However, achieving reasonable performance for these systems is difficult for PID controllers. Therefore, the PID controller combines with nonlinear gain called the NPID controller, that is designed to control the EHSAS position tracking. The nonlinear gain is used to enhance system efficiency and minimize overshoot by using a relatively higher gain~\cite{b21}. The Equation \ref{npid} represents the NPID controller.

\begin{figure}[t]
	\centering
	\resizebox{9cm}{!}{\includegraphics{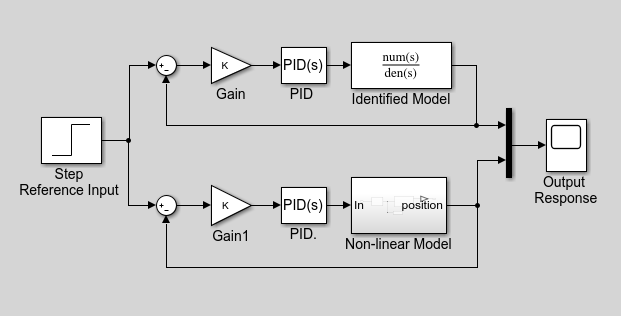}}	
	\caption{Simulink model with NPID controller.}
	\label{PID model}
\end{figure}

\begin{equation}
u(t) = K_p k(e) e(t) + K_i k(e) \underset{0}{\overset{t}{\int }} e(t) dt + K_d k(e) \frac{d}{dt} e(t)
\label{npid}
\end{equation}
A wide variety of options are available for the nonlinear gain $k$. Here, as a function of error $e$, nonlinear gain $k$ has been used as the hyperbolic function~\cite{b21}.
\begin{equation}
k= k_{0}+k_{1}\left[{1-sech(k_{2}e)}\right]
\label{npid gain}
\end{equation}
\begin{equation}
k= k_{0}+k_{1}\left[{1-\dfrac{2}{\exp(k_{2}e)+exp(-k_{2}e)}}\right]
\label{npid2 gain}
\end{equation}
\begin{align*}
k_{max} &= k_{0}+k_{1} \ ; \: \: \: \: \: \: \: \: \: \: e=\pm \infty \\
k_{min} &= k_{0} \ ; \: \: \: \: \: \: \: \: \: \: \: \: \: \: \: \:\: \: \: \: e=0
\end{align*}
where $k_{0}$, $k_{1}$, $k_{2}$ represents the constant values fed by the user, that has been selected through experimentation. However, $k_{0}$ shows the minimum value, the range of variation defined by $k_{1}$, and the rate of variation of $k$ specified by $k_{2}$. The experiment is repeated with $k$ as a hyperbolic function of $e$, with the effect of the nonlinear gain $k$ on the system performance, as shown

\begin{equation}
k= 4-3sech(0.05e)
\label{npid gain3}
\end{equation}

The nonlinear gain $k$ varies with respect to the error. The gain is automatically minimized as the time continues and the error is reduced, then eventually settles to the final value of one, with zero steady state error.
%
			
			
			

\begin{figure*} [!t]
	\centering
	\begin{subfigure} [t]{0.48\textwidth}
		\includegraphics[width=1\textwidth]{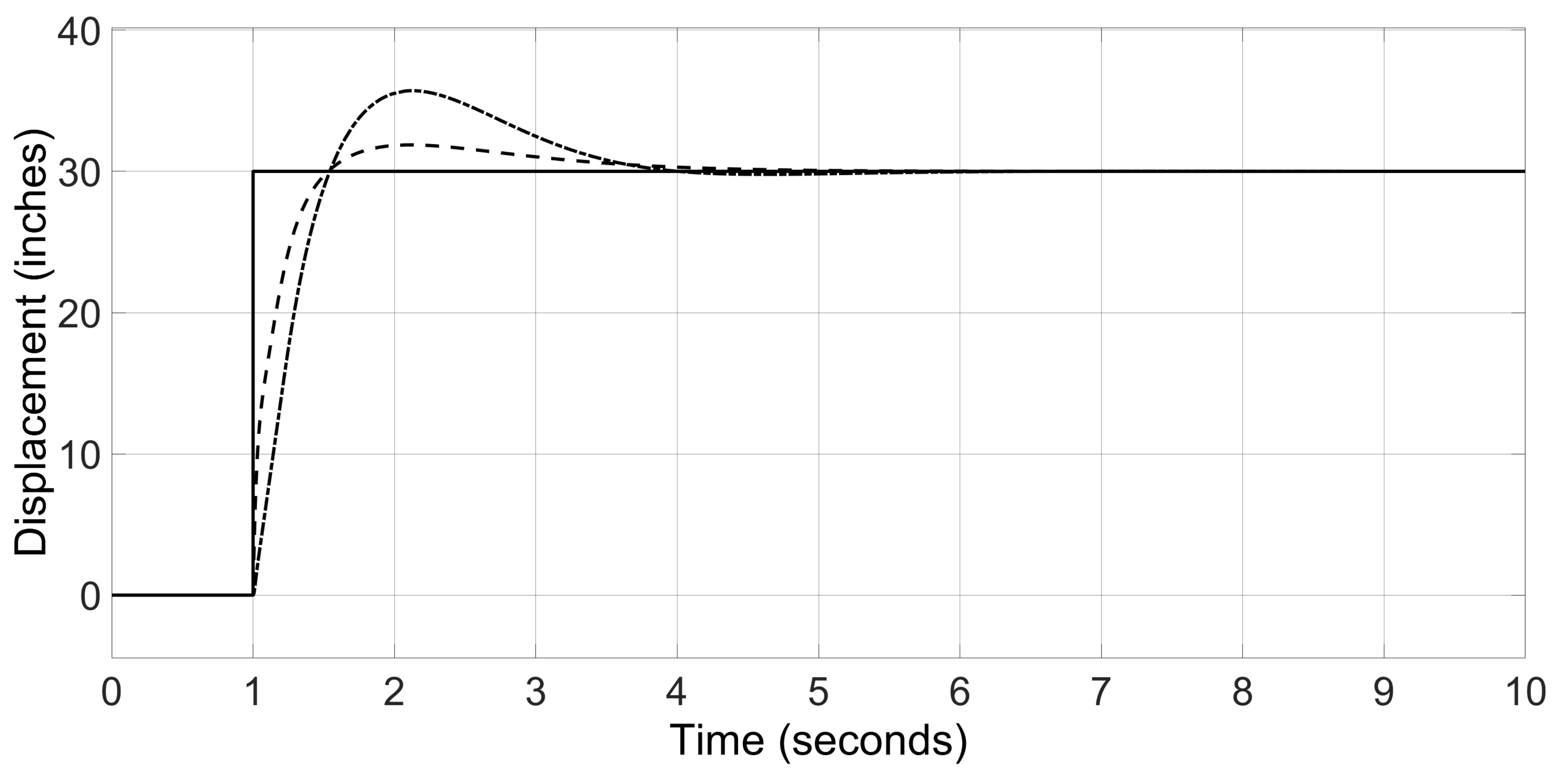}  
		\caption{Step Response Comparison for Chirp Identified Model}
		\label{fig:sub-first}
	\end{subfigure}
	\begin{subfigure}[t]{0.48\textwidth}
		\includegraphics[width=1\textwidth]{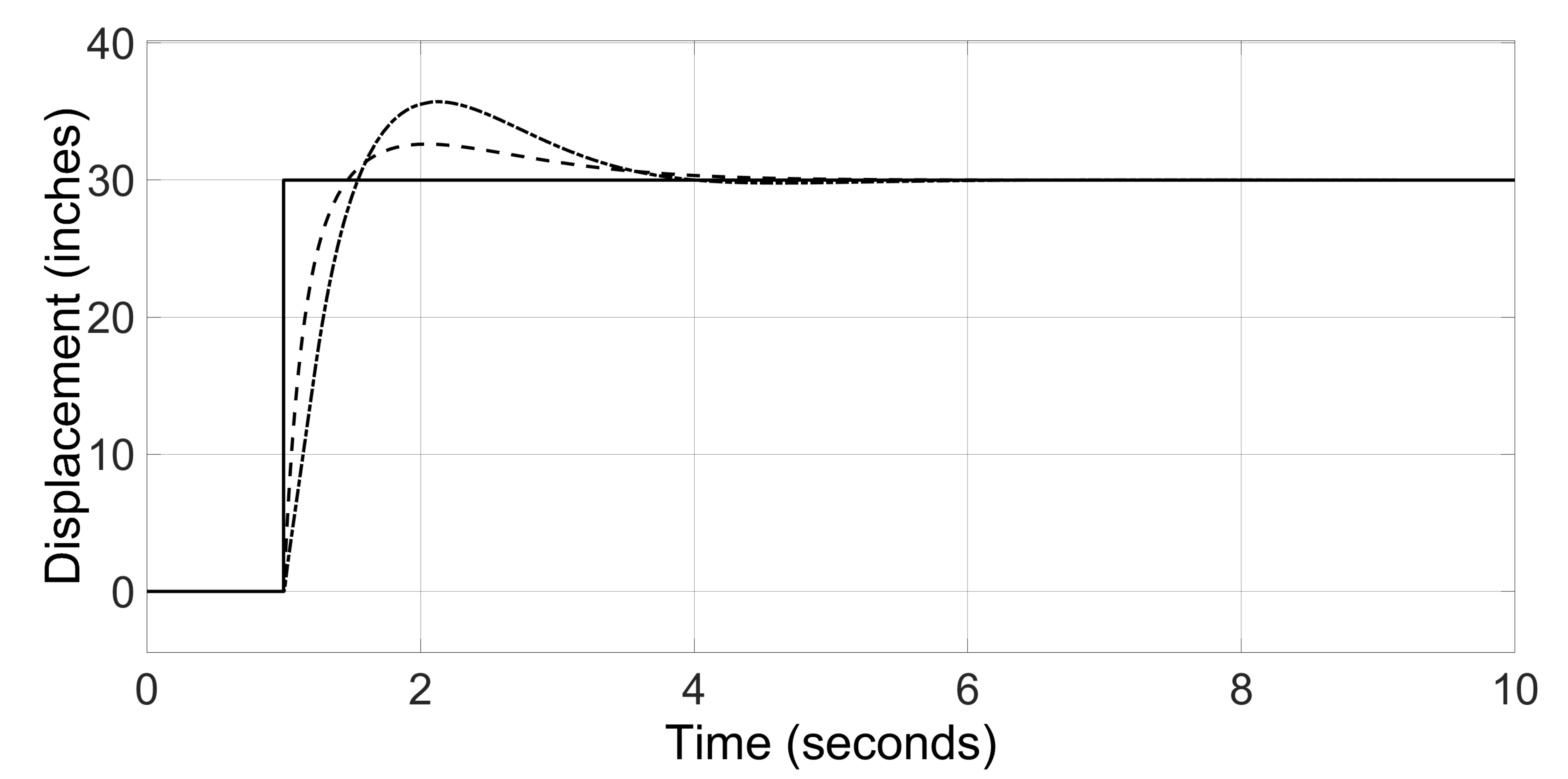}  
		\caption{Step Response Comparison for Multisine Identified Model}
		\label{fig:sub-second}
	\end{subfigure}
	
	\begin{subfigure}[t]{0.48\textwidth}
		\includegraphics[width=1\textwidth]{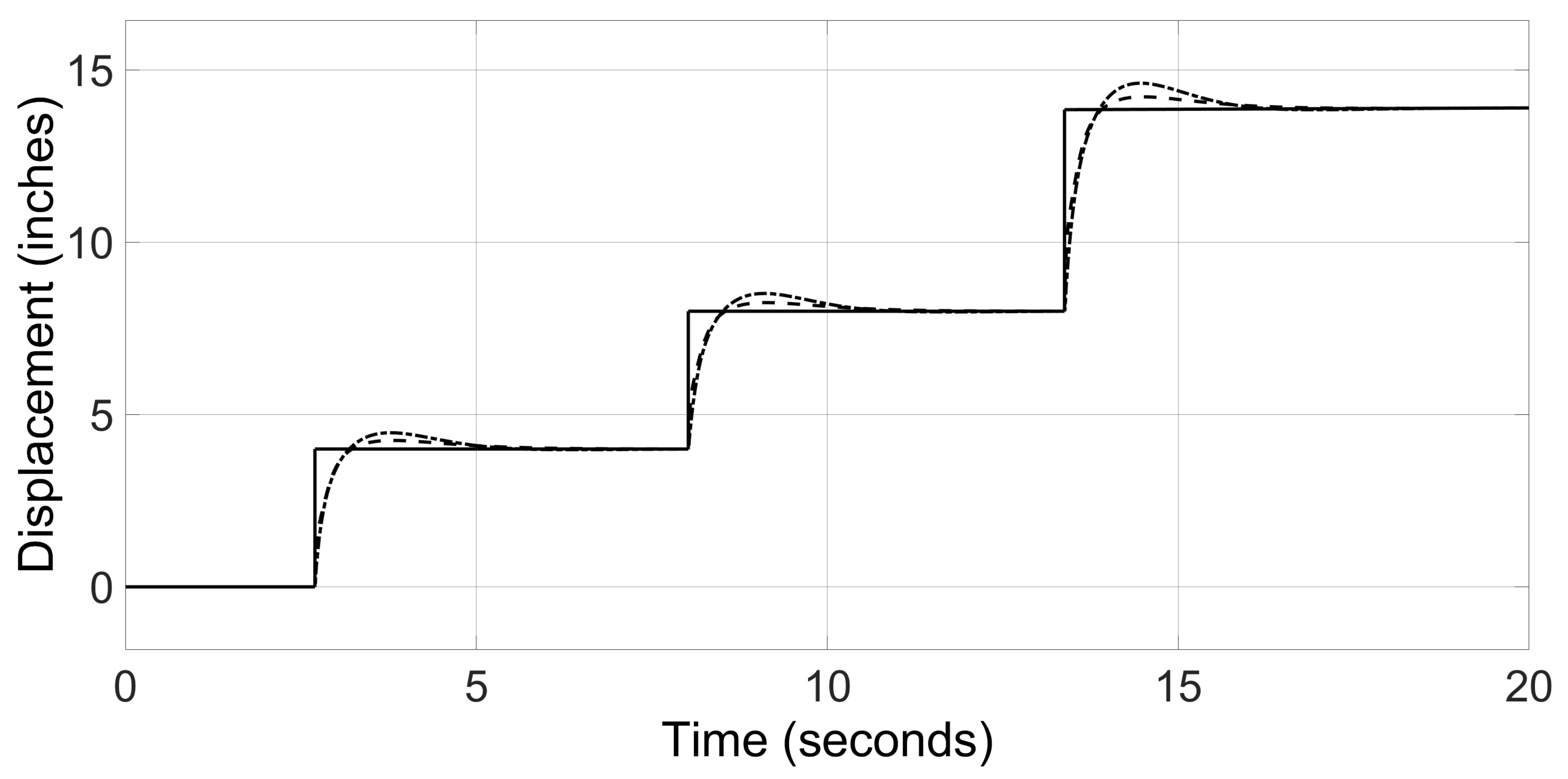}  
		\caption{Staircase Response Comparison for Chirp Identified Model}
		\label{fig:sub-third}
	\end{subfigure}
	\begin{subfigure}[t]{0.48\textwidth}
		\includegraphics[width=1\textwidth]{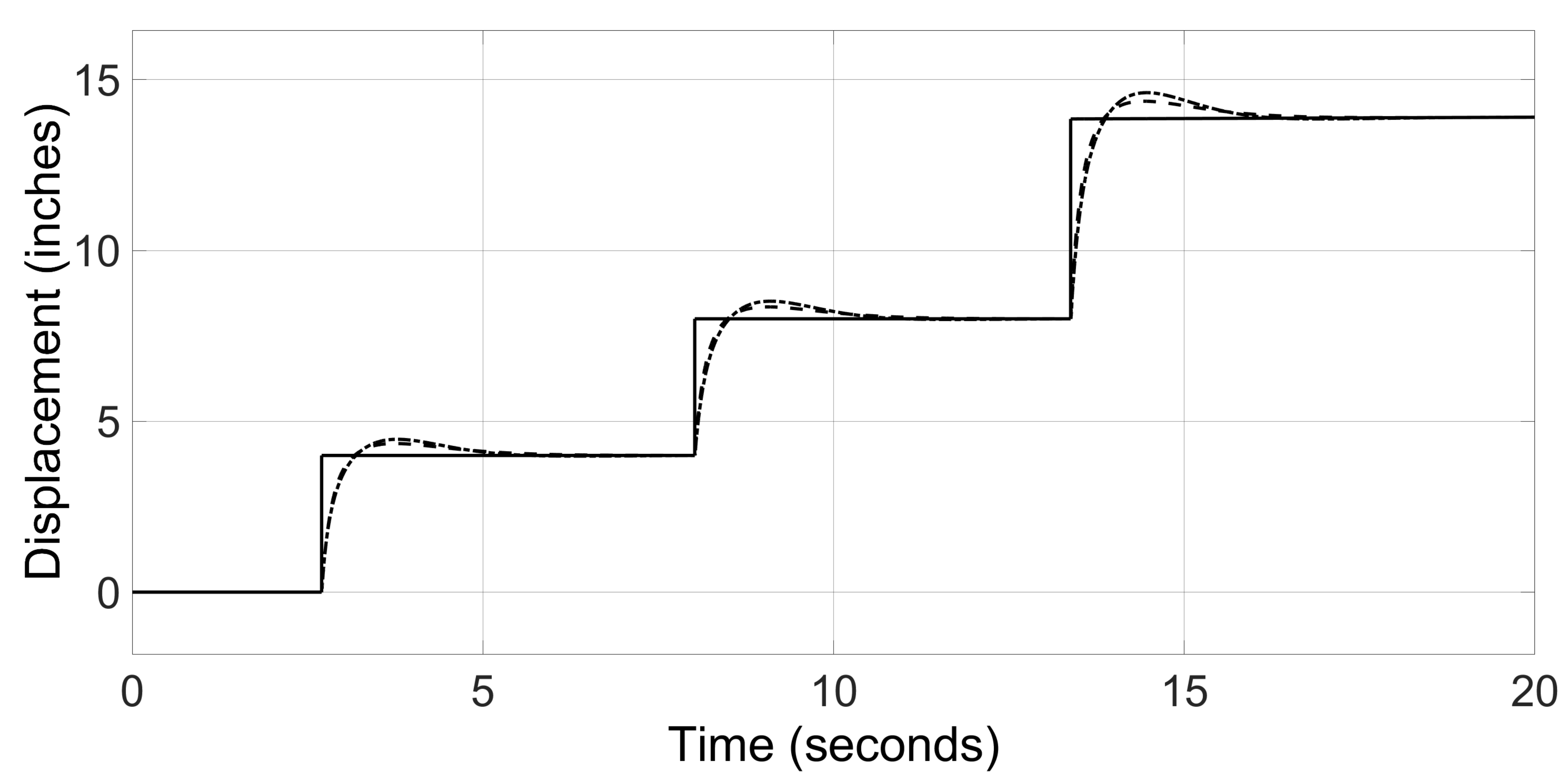}  
		\caption{Staircase Response Comparison for Multisine Identified Model}
		\label{fig:sub-fourth}
	\end{subfigure}
	
	\begin{subfigure}[t]{0.48\textwidth}
		\includegraphics[width=1\textwidth]{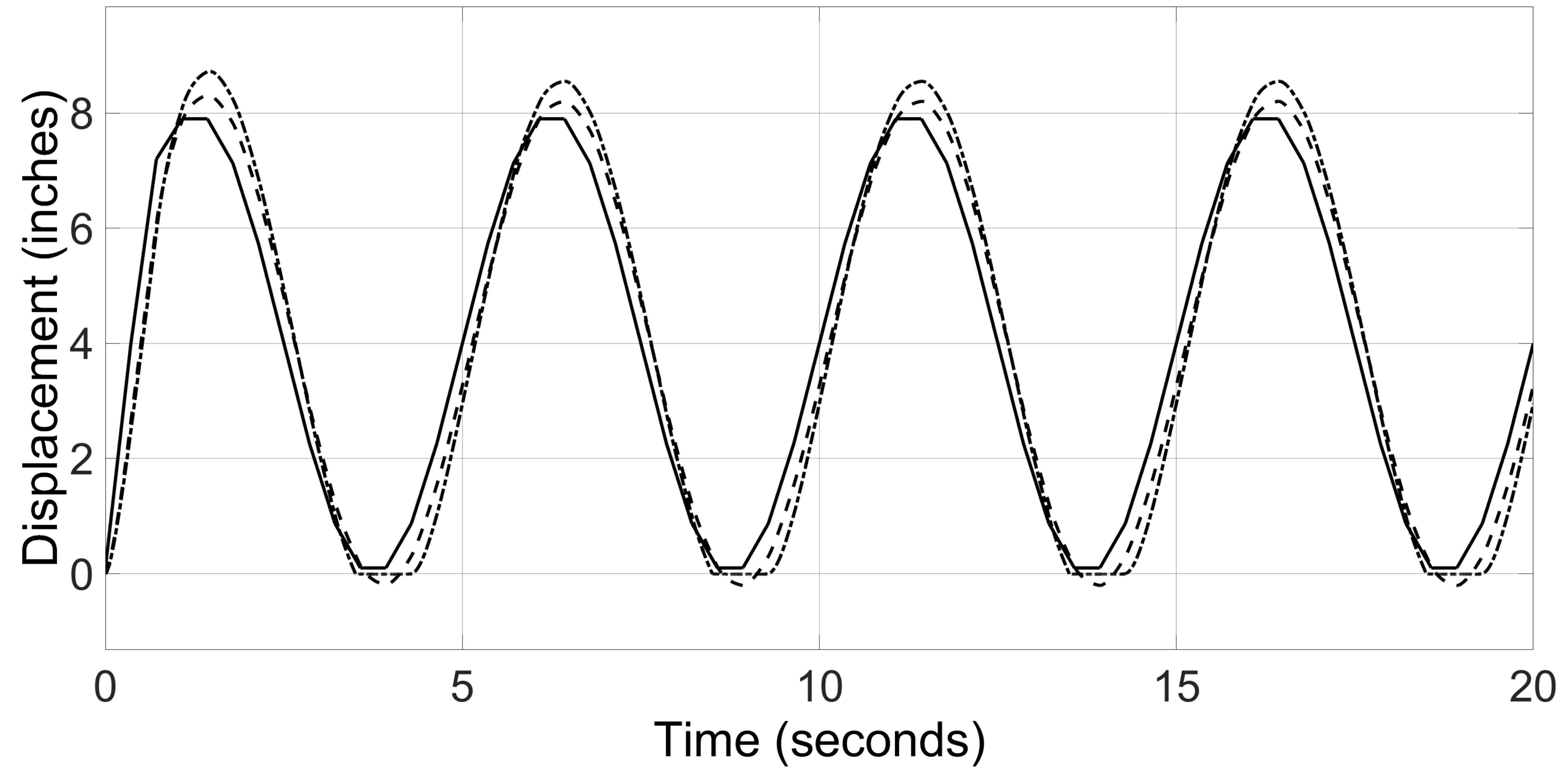}  
		\caption{Sinusoidal Response Comparison for Chirp Identified Model}
		\label{fig:sub-fifth}
	\end{subfigure}
	\begin{subfigure}[t]{0.48\textwidth}
		\includegraphics[width=1\textwidth]{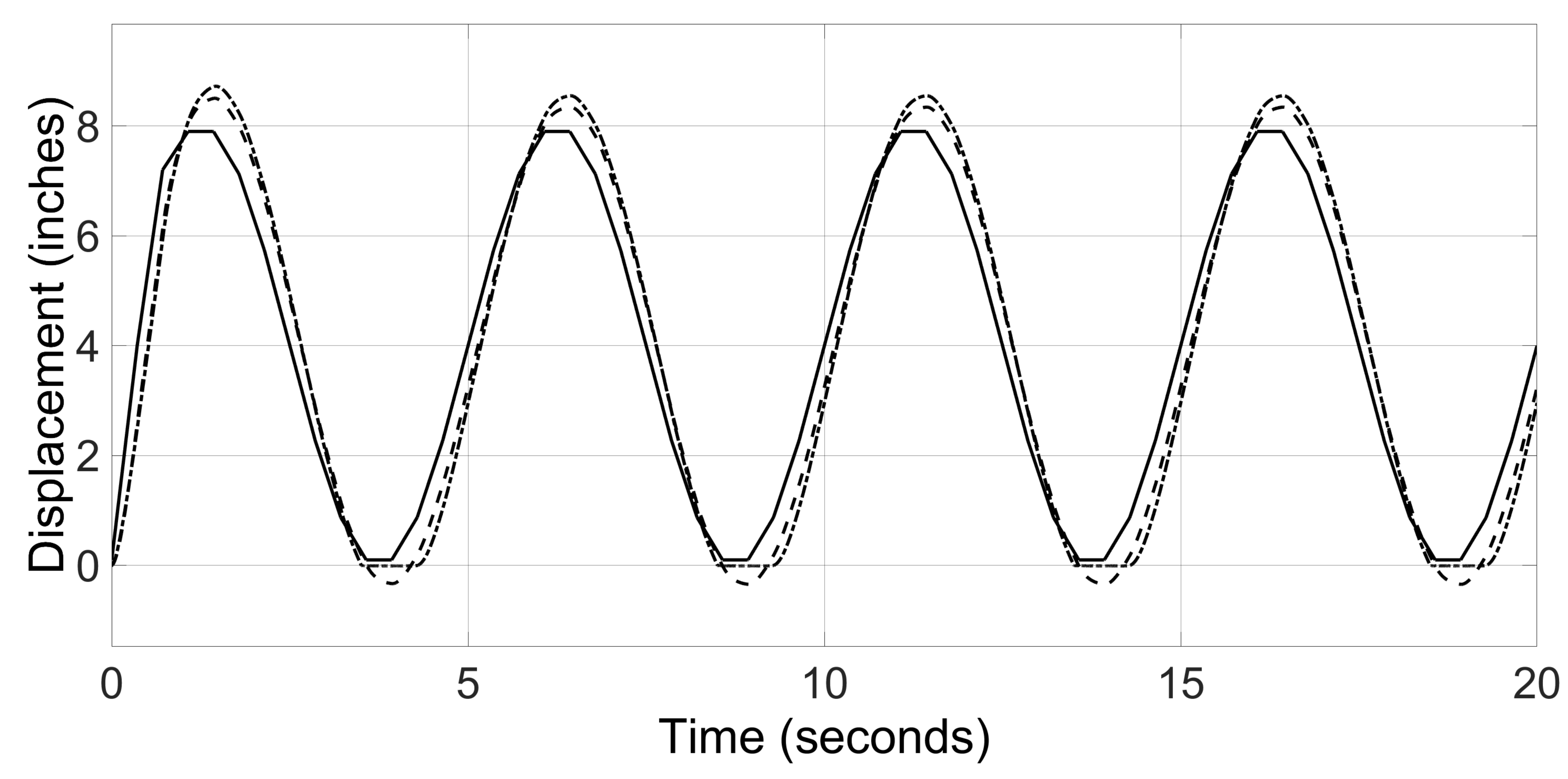}  
		\caption{Sinusoidal Response Comparison for Multisine Identified Model}
		\label{fig:sub-sixth}
	\end{subfigure}
	\caption{Comparative results of controller performance between Identified Model and Nonlinear Model by using different reference signals (Left hand side shows the obtained model through chirp excitation signal whereas, right hand side shows the obtained model through multisine excitation signal).   \\ \hdashrule{3cm}{0.5pt}{} represents the reference signal,
    \hdashrule[0.5ex]{3cm}{0.3mm}{3mm 3pt 1mm 5pt} shows the nonlinear model,  \hdashrule[0.5ex]{3cm}{0.3mm}{3mm} shows an identified model. }
	\label{testing_results}
	
\end{figure*}
 Although the output of the device needs to achieve zero steady state error with this controller, proportional controller $K_p$ is used to improve the speed of response so that it can track the position of hydraulic actuator. The system is given integral controller $K_i$ in order to get the steady state error zero or very low. Derivative controller  $K_d$ will improve system speed performance~\cite{b11}. The derivative action may not be appropriate at times because the proportional and integral action already provides a reasonable response to the output. Ziegler-Nichols tuning method is used to determine the tuning value of $K_p$, $K_i$, and $K_d$. Before the tuning process, the critical gain, $K_{cr}$ and critical oscillation period, $T_{cr}$ needs to be determined. Based on these two parameters, the value of $K_p$, $K_i$, and $K_d$ are adjusted. Ziegler-Nichols tuning rules and these values might be modified to obtain the best output response. 
 
The nonlinear PID controller has been applied to improve the position performance of the system. Fig. \ref{PID model} illustrates the Simulink model with composed of NPID controller. The tracking performance of controller has been verified by feeding different types of reference signals like step, staircase, and sine. These reference signals have been given to both; the nonlinear model and the identified model for position tracking capability analysis. Fig. \ref{testing_results} clearly shows that both inherited output signals from the chirp and multisine excitation signal followed the nearest path to the reference signal. The performance of the controller has been investigated with the transient response analysis, including the rise time $T_r$, settling time $T_s$, and over-shoot $OS$, as presented in Table. \ref{PID Characteristics}.

\section{Conclusion}
In this research, selection of suitable excitation signals for system identification of hydraulic actuator has been presented. A proposed methodology for selection of excitation signal parameter in relation to actuator dynamics has been validated in simulation by comparing performance with a nonlinear model of the actuator. After that, a nonlinear PID controller is configured to control the electro-hydraulic servo actuator system, based on the model acquired from the identification process. The position tracking capability has been evaluated through different reference signals. Simulation results show an accurate tracking performance. Future work includes experimental verification of the proposed techniques and implementation on a hydraulic actuator.

\begin{center}
	\begin{table} [ht]
		\centering
		\caption{Transient response analysis.}
		\vspace{1ex}
		\label{PID Characteristics}
		\begin{tabular}{|l|l|l|l|l|}

			 \hline 
		
			\multicolumn{2}{|c|}{\textbf{Model}} & $T_r$ ($s$)   &  $T_s$ ($s$)  & $OS$ ($\%$)      \\     \hline
			
		\begin{tabular}{c}
			\textbf{Model 1} \\ \textit{Chirp} \\
		\end{tabular}
			 & 
			 	\begin{tabular}{c}
			 Identified Model \\ Nonlinear Model \\
			\end{tabular} 
		&
		\begin{tabular}{c}
		 0.25 \\ 0.28 \\
		 	\end{tabular} 
	 	&
	 	\begin{tabular}{c}
	 	2.1 \\ 2.2 \\
	 	\end{tabular} 
 	&		 
		 \begin{tabular}{c}
		 	 4.5 \\ 4.8 \\
		 \end{tabular} 
	 \\ \hline
	 
	 \begin{tabular}{c}
	 	\textbf{Model 2} \\ \textit{Multisine} \\
	 \end{tabular}
	 & 
	 \begin{tabular}{c}
	 	Identified Model \\ Nonlinear Model 
	 \end{tabular} 
 &	 
	 \begin{tabular}{c}
	 	0.1 \\ 0.3 \\
	 \end{tabular}
  &	 
	 \begin{tabular}{c}
	 	2.8 \\ 3.0 \\
	 \end{tabular} 
 &	 
	 \begin{tabular}{c}
	 	3.2 \\ 4.3 \\
	 \end{tabular} 
 \\ \hline
		\end{tabular}
	\end{table}
\end{center}

\balance
\bibliographystyle{IEEEtran}
\bibliography{zainab}

\begin{thebibliography}{10}
\providecommand{\url}[1]{#1}
\csname url@samestyle\endcsname
\providecommand{\newblock}{\relax}
\providecommand{\bibinfo}[2]{#2}
\providecommand{\BIBentrySTDinterwordspacing}{\spaceskip=0pt\relax}
\providecommand{\BIBentryALTinterwordstretchfactor}{4}
\providecommand{\BIBentryALTinterwordspacing}{\spaceskip=\fontdimen2\font plus
\BIBentryALTinterwordstretchfactor\fontdimen3\font minus
  \fontdimen4\font\relax}
\providecommand{\BIBforeignlanguage}[2]{{%
\expandafter\ifx\csname l@#1\endcsname\relax
\typeout{** WARNING: IEEEtran.bst: No hyphenation pattern has been}%
\typeout{** loaded for the language `#1'. Using the pattern for}%
\typeout{** the default language instead.}%
\else
\language=\csname l@#1\endcsname
\fi
#2}}
\providecommand{\BIBdecl}{\relax}
\BIBdecl

\bibitem{b1}
J.~Choi, ``Robust position control of electro-hydrostatic actuator systems with
  radial basis function neural networks,'' \emph{Journal of Advanced Mechanical
  Design, Systems, and Manufacturing}, vol.~7, no.~2, pp. 257--267, 2013.

\bibitem{b45}
O.~Nelles, \emph{Nonlinear system identification: from classical approaches to
  neural networks and fuzzy models}.\hskip 1em plus 0.5em minus 0.4em\relax
  Springer Science \& Business Media, 2013.

\bibitem{b46}
I.~Koblen and J.~Kov{\'a}cov{\'a}, ``Selected information on flight
  simulators-main requirements, categories and their development, production
  and using for flight crew training in the both slovak republic and czech
  republic conditions,'' \emph{Incas Bulletin}, vol.~4, no.~3, p.~73, 2012.

\bibitem{b47}
\BIBentryALTinterwordspacing
``Electrohydraulic,'' East Aurora, NY, August 2013. [Online]. Available:
  \url{www.moog.com/products/actuators-servoactuators/actuation-technologies}
\BIBentrySTDinterwordspacing

\bibitem{b3}
M.~Kalyoncu and M.~Haydim, ``Mathematical modelling and fuzzy logic based
  position control of an electrohydraulic servosystem with internal leakage,''
  \emph{Mechatronics}, vol.~19, no.~6, pp. 847--858, 2009.

\bibitem{b2}
Z.~Wang, J.~Shao, J.~Lin, and G.~Han, ``Research on controller design and
  simulation of electro-hydraulic servo system,'' in \emph{2009 International
  Conference on Mechatronics and Automation}.\hskip 1em plus 0.5em minus
  0.4em\relax IEEE, 2009, pp. 380--385.

\bibitem{b4}
M.~F. Rahmat, S.~Md~Rozali, N.~Abdul~Wahab, Z.~Has, and K.~Jusoff, ``Modeling
  and controller design of an electro-hydraulic actuator system,''
  \emph{American Journal of Applied Sciences}, vol.~7, no.~8, pp. 1100--1108,
  2010.

\bibitem{b10}
N.~Izzuddin, M.~R. Johari, K.~Osman \emph{et~al.}, ``System identification and
  predictive functional control for electro-hydraulic actuator system,'' in
  \emph{2015 IEEE International Symposium on Robotics and Intelligent Sensors
  (IRIS)}.\hskip 1em plus 0.5em minus 0.4em\relax IEEE, 2015, pp. 138--143.

\bibitem{b8}
N.~Ishak, M.~Tajjudin, H.~Ismail, and R.~Adnan, ``System identification and
  model validation of electro-hydraulic actuator for quarter car system,''
  2006.

\bibitem{b16}
G.~Ren, J.~Song, and N.~Sepehri, ``Design of a low-bandwidth position
  controller based on system identification for an electro-hydrostatic
  actuator,'' \emph{Proceedings of the Institution of Mechanical Engineers,
  Part I: Journal of Systems and Control Engineering}, vol. 232, no.~2, pp.
  149--160, 2018.

\bibitem{b7}
R.~Ghazali, C.~Soon, H.~Jaafar, Y.~M. Sam, and M.~Rahmat, ``System
  identification of electro-hydraulic actuator system with pressure and load
  effects,'' in \emph{2014 IEEE International Conference on Control System,
  Computing and Engineering (ICCSCE 2014)}.\hskip 1em plus 0.5em minus
  0.4em\relax IEEE, 2014, pp. 256--260.

\bibitem{b17}
S.~Mohammed, C.~C. Soon, R.~Ghazali, A.~A. Yusof, Y.~M. Sam, and C.~M. Shern,
  ``An electro-hydraulic servo with intelligent control strategy,'' in
  \emph{MATEC Web of Conferences}, vol. 150.\hskip 1em plus 0.5em minus
  0.4em\relax EDP Sciences, 2018, p. 01016.

\bibitem{b18}
X.~W. Liang, Z.~H. Ismail \emph{et~al.}, ``System identification and model
  predictive control using cvxgen for electro-hydraulic actuator,''
  \emph{International Journal of Integrated Engineering}, vol.~11, no.~4, 2019.

\bibitem{b5}
L.~Li and T.~Thurner, ``Accurate modeling and identification of servo-hydraulic
  cylinder systems in multi-axial test applications,'' \emph{Ventil}, vol.~19,
  no.~6, pp. 462--470, 2013.

\bibitem{b15}
L.~Jin and Q.~Wang, ``Accurate model identification of the inertial mass
  dynamic of hydraulic cylinder with model uncertainty,'' \emph{Proceedings of
  the Institution of Mechanical Engineers, Part I: Journal of Systems and
  Control Engineering}, vol. 233, no.~5, pp. 501--510, 2019.

\bibitem{b9}
W.-f. ZHONG and X.-x. HE, ``Fuzzy neural network control of electro-hydraulic
  position servo system [j],'' \emph{Electric Machines and Control}, vol.~4,
  2008.

\bibitem{b6}
D.~M. Wonohadidjojo, G.~Kothapalli, and M.~Y. Hassan, ``Position control of
  electro-hydraulic actuator system using fuzzy logic controller optimized by
  particle swarm optimization,'' \emph{International Journal of Automation and
  Computing}, vol.~10, no.~3, pp. 181--193, 2013.

\bibitem{b11}
S.~M. Rozali, M.~F. Rahmat, N.~A. Wahab, R.~Ghazali \emph{et~al.}, ``Pid
  controller design for an industrial hydraulic actuator with servo system,''
  in \emph{2010 IEEE Student Conference on Research and Development
  (SCOReD)}.\hskip 1em plus 0.5em minus 0.4em\relax IEEE, 2010, pp. 218--223.

\bibitem{b12}
J.~Richalet and D.~O'Donovan, ``Elementary predictive functional control: A
  tutorial,'' in \emph{2011 International Symposium on Advanced Control of
  Industrial Processes (ADCONIP)}.\hskip 1em plus 0.5em minus 0.4em\relax IEEE,
  2011, pp. 306--313.

\bibitem{b13}
S.~Cho and K.~Edge, ``Adaptive sliding mode tracking control of hydraulic
  servosystems with unknown non-linear friction and modelling error,''
  \emph{Proceedings of the Institution of Mechanical Engineers, Part I: Journal
  of Systems and Control Engineering}, vol. 214, no.~4, pp. 247--257, 2000.

\bibitem{b14}
R.~Ghazali, ``Adaptive discrete sliding mode control of an electro-hydraulic
  actuator system,'' Ph.D. dissertation, Universiti Teknologi Malaysia, 2013.

\bibitem{b40}
C.~M. Shern, R.~Ghazali, C.~S. Horng, H.~I. Jaafar, and C.~C. Soon,
  ``Performance analysis of position tracking control with pid controller using
  an improved optimization technique,'' \emph{Evolution (DE)}, vol.~11, p.~12,
  2019.

\bibitem{b50}
J.~Hasan, T.~Karmaker, and M.~I. Ahmed, ``Mathematical modeling and simulation
  based system identification of non-minimum phase electro-hydraulic servo
  (ehs) system.''

\bibitem{b48}
\BIBentryALTinterwordspacing
``Actuators-servoactuators,'' East Aurora, NY, 2013. [Online]. Available:
  \url{www.moog.com/products/actuators-servoactuators}
\BIBentrySTDinterwordspacing

\bibitem{b19}
L.~Ljung, \emph{\textbf{Practical Issues} of System Identification}.\hskip 1em
  plus 0.5em minus 0.4em\relax Link{\"o}ping University Electronic Press, 2007.

\bibitem{b49}
M.~Barenthin, ``On input design in system identification for control,'' Ph.D.
  dissertation, Signaler, sensorer och system, 2006.

\bibitem{b44}
P.~Flandrin, ``Chirps everywhere,'' \emph{CNRS-Lyon}, 2002.

\bibitem{b20}
S.~Carnduff, ``Aircraft and rotorcraft system identification: Engineering
  methods with flight test examples mb tischler and rk remple american
  institute of aeronautics and astronautics, 1801 alexander bell drive, suite
  500, reston, va 20191-4344, usa. 2006. 523pp. illustrated. $83.95 (aiaa
  members), $119.95 (non-members). isbn 1-56347-837-4.'' \emph{The Aeronautical
  Journal}, vol. 111, no. 1123, pp. 601--601, 2007.

\bibitem{b23}
T.~Ling, M.~Rahmat, A.~Husain, and R.~Ghazali, ``System identification of
  electro-hydraulic actuator servo system,'' in \emph{2011 4th International
  Conference on Mechatronics (ICOM)}.\hskip 1em plus 0.5em minus 0.4em\relax
  IEEE, 2011, pp. 1--7.

\bibitem{b30}
X.~Zuo, J.-w. Liu, X.~Wang, and H.-q. Liang, ``Adaptive pid and model reference
  adaptive control switch controller for nonlinear hydraulic actuator,''
  \emph{Mathematical Problems in Engineering}, vol. 2017, 2017.

\bibitem{b21}
H.~Seraji, ``A new class of nonlinear pid controllers with robotic
  applications,'' \emph{Journal of Robotic Systems}, vol.~15, no.~3, pp.
  161--181, 1998.

\bibitem{b22}
C.~W. De~Silva, \emph{Control sensors and actuators}.\hskip 1em plus 0.5em
  minus 0.4em\relax Prentice Hall PTR, 1988.

\bibitem{b24}
R.~Ghazali, Y.~Sam, M.~Rahmat, and A.~Hashim, ``Zulfatman," sliding mode
  control with pid sliding surface of an electro-hydraulic servo system for
  position tracking control,",'' \emph{Australian Journal of Basic and Applied
  Sciences}, vol.~4, no.~10, pp. 4749--4759, 2010.

\bibitem{b25}
B.~Li, J.~Yan, G.~Guo, H.~Wang, and M.~Zhang, ``Identification of dynamic
  parameters and friction coefficients for a heavy-duty hydraulic
  manipulator,'' in \emph{Proceedings of the 10th World Congress on Intelligent
  Control and Automation}.\hskip 1em plus 0.5em minus 0.4em\relax IEEE, 2012,
  pp. 3102--3106.

\bibitem{b26}
C.~C. Maier, S.~Schr{\"o}ders, W.~Ebner, M.~K{\"o}ster, A.~Fidlin, and
  C.~Hametner, ``Modeling and nonlinear parameter identification for hydraulic
  servo-systems with switching properties,'' \emph{Mechatronics}, vol.~61, pp.
  83--95, 2019.

\bibitem{b27}
A.~Novak, L.~Simon, F.~Kadlec, and P.~Lotton, ``Nonlinear system identification
  using exponential swept-sine signal,'' \emph{IEEE Transactions on
  Instrumentation and Measurement}, vol.~59, no.~8, pp. 2220--2229, 2009.

\bibitem{b28}
T.~S. Xavier, E.~C.~G. Barbosa, and L.~C. G{\'o}es, ``Predictive control design
  based on system identification of an electro-hydraulic actuator applied to
  brazilian sounding rockets and microsatellite launchers,'' in
  \emph{WIEFP2018--4th Workshop on Innovative Engineering for Fluid Power,
  November 28-30, Sao Paulo, Brazil}, no. 156.\hskip 1em plus 0.5em minus
  0.4em\relax Link{\"o}ping University Electronic Press, 2018, pp. 63--68.

\bibitem{b29}
R.~{Adnan}, M.~{Tajjudin}, N.~{Ishak}, H.~{Ismail}, and M.~H. {Fazalul
  Rahiman}, ``Self-tuning fuzzy pid controller for electro-hydraulic
  cylinder,'' in \emph{2011 IEEE 7th International Colloquium on Signal
  Processing and its Applications}, March 2011, pp. 395--398.

\bibitem{b31}
M.~Ahmed, B.~D. Halilu, M.~Idi, and B.~Mohammed, ``System identification and
  control of a hydraulic actuator,'' 2006.

\bibitem{b32}
K.-E. Rydberg, ``Hydraulic servo systems: Dynamic properties and control,''
  2016.

\bibitem{b33}
P.~Van~den Hof, X.~Bombois, and L.~N.~D. Course, ``System identification for
  control,'' \emph{Delft Center for Systems and Control, TU-Delft. Lecture
  notes, Dutch Institute for Systems and Control (DISC)}, 2004.

\bibitem{b34}
G.~K. Mohan, M.~Naidu, N.~M. Rao \emph{et~al.}, ``An implementation of
  different non linear pid controllers on a single tank level control using
  matlab,'' \emph{International Journal of Computer Applications}, vol.~54,
  no.~1, 2012.

\bibitem{b35}
G.~Zaidner, S.~Korotkin, E.~Shteimberg, A.~Ellenbogen, M.~Arad, and Y.~Cohen,
  ``Non linear pid and its application in process control,'' in \emph{2010 IEEE
  26-th Convention of Electrical and Electronics Engineers in Israel}.\hskip
  1em plus 0.5em minus 0.4em\relax IEEE, 2010, pp. 000\,574--000\,577.

\bibitem{b36}
Z.~Wang, J.~Shao, J.~Lin, and G.~Han, ``Research on controller design and
  simulation of electro-hydraulic servo system,'' in \emph{2009 International
  Conference on Mechatronics and Automation}.\hskip 1em plus 0.5em minus
  0.4em\relax IEEE, 2009, pp. 380--385.

\bibitem{b37}
J.~Hasan, T.~Karmaker, and M.~I. Ahmed, ``Mathematical modeling and simulation
  based system identification of non-minimum phase electro-hydraulic servo
  (ehs) system.''

\bibitem{b38}
M.~Pencelli, R.~Villa, A.~Argiolas, G.~Ferretti, M.~Niccolini, M.~Ragaglia,
  P.~Rocco, and A.~M. Zanchettin, ``Accurate dynamic modelling of hydraulic
  servomechanisms,'' in \emph{2019 Design, Automation \& Test in Europe
  Conference \& Exhibition (DATE)}.\hskip 1em plus 0.5em minus 0.4em\relax
  IEEE, 2019, pp. 1257--1260.

\bibitem{b39}
C.~C. Maier, S.~Schr{\"o}ders, W.~Ebner, M.~K{\"o}ster, A.~Fidlin, and
  C.~Hametner, ``Modeling and nonlinear parameter identification for hydraulic
  servo-systems with switching properties,'' \emph{Mechatronics}, vol.~61, pp.
  83--95, 2019.

\bibitem{b42}
P.~Annus, R.~Land, M.~Min, and J.~Ojarand, ``Simple signals for system
  identification,'' \emph{Fourier Transform-Signal Processing. InTech: Rijeka},
  pp. 257--276, 2012.

\bibitem{b43}
J.~Heeg and E.~Morelli, ``Evaluation of simultaneous multisine excitation of
  the joined wing aeroelastic wind tunnel model,'' in \emph{52nd
  AIAA/ASME/ASCE/AHS/ASC Structures, Structural Dynamics and Materials
  Conference 19th AIAA/ASME/AHS Adaptive Structures Conference 13t}, 2011, p.
  1959.

\end{thebibliography}

\vspace{17pt}

\newpage

\begin{IEEEbiography}[{\includegraphics[width=1in,height=1.25in,clip,keepaspectratio]{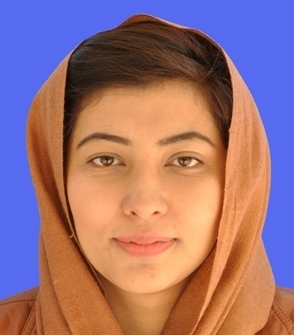}}]{Zainab Nisar} received her BS degree in Electrical Engineering in 2017. She is currently pursuing her MS degree in Avionics Engineering from the College of Aeronautical Engineering, National University of Sciences and Technology (NUST), Pakistan. Her research interests include system identification, feedback control, image processing and steganography.
\end{IEEEbiography}

\begin{IEEEbiography}[{\includegraphics[width=1in,height=1.25in,clip,keepaspectratio]{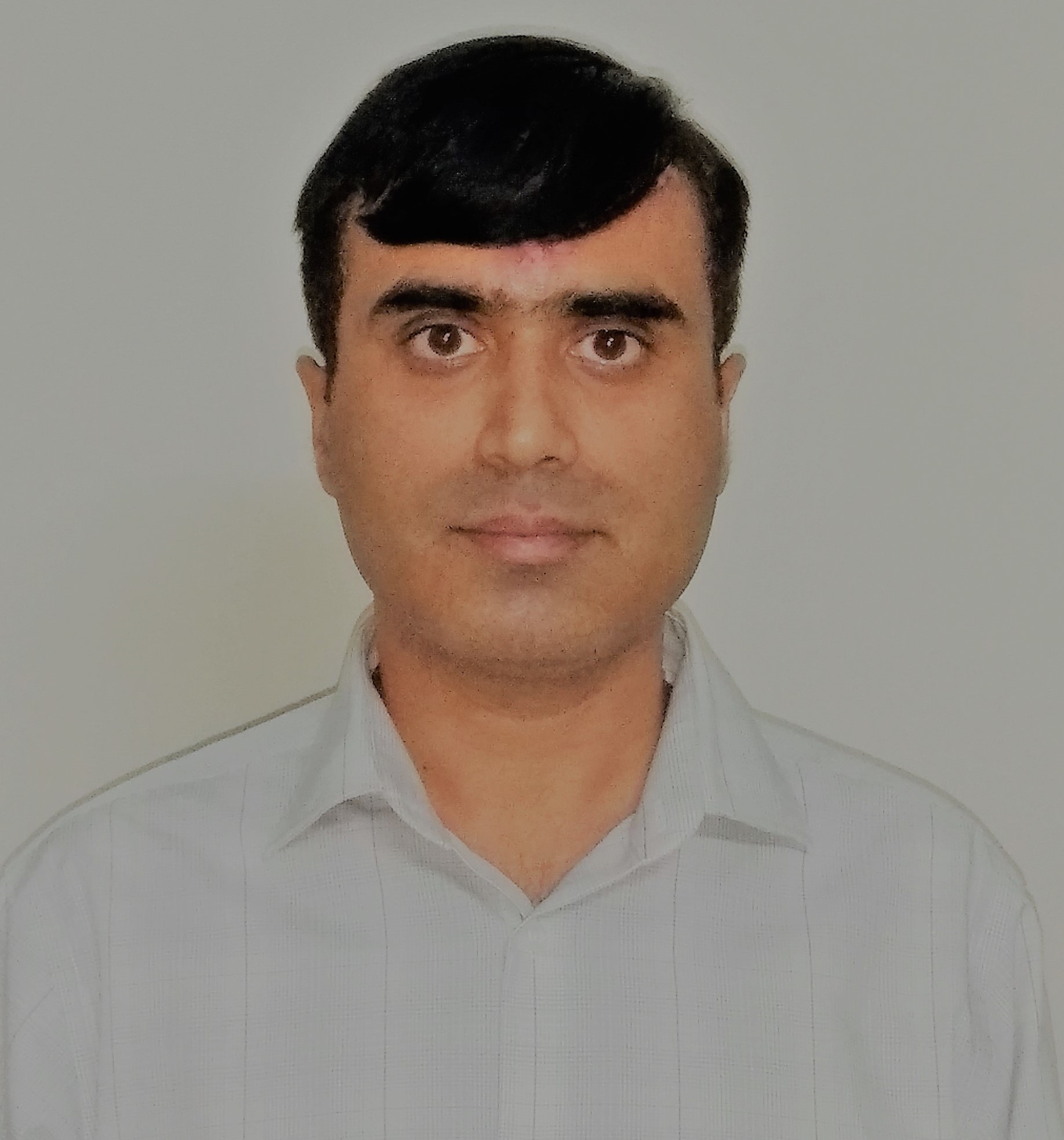}}]{Dr Hammad Munawar} received his PhD degree in 2017 and MS degree in 2012. He is currently an Assistant Professor at the Department of Avionics Engineering, College of Aeronautical Engineering, National University of Sciences and Technology (NUST), Pakistan. His research interests include feedback control, haptics and force control.
\end{IEEEbiography}

\EOD

\end{document}